# Nepotistic Relationships in Twitter and their Impact on Rank Prestige Algorithms

Daniel Gayo-Avello[1]

Department of Computer Science, University of Oviedo, SPAIN

**Abstract.** Micro-blogging services such as Twitter allow anyone to publish anything, anytime. Needless to say, many of the available contents can be diminished as babble or spam. However, given the number and diversity of users, some valuable pieces of information should arise from the stream of tweets. Thus, such services can develop into valuable sources of up-to-date information (the so-called real-time web) provided a way to find the most relevant/trustworthy/authoritative users is available. Hence, this makes a highly pertinent question for which graph centrality methods can provide an answer. In this paper the author offers a comprehensive survey of feasible algorithms for ranking users in social networks, he examines their vulnerabilities to linking malpractice in such networks, and suggests an objective criterion against which to compare such algorithms. Additionally, he suggests a first step towards "desensitizing" prestige algorithms against cheating by spammers and other abusive users.

# 1. Introduction

Twitter is a service which allows users to publish short text messages (tweets) which are shown to other users following the author of the message. In case the author is not protecting his tweets, they appear in the so-called public timeline and they are served as search results in response to user submitted queries. Thus, Twitter can be a source of valuable real-time information and, in fact, several major search engines are including tweets as search results.

Given that tweets are published by individual users, ranking them to find the most relevant information is a crucial matter. Indeed, at the moment of this writing, Google seems to be already applying the *PageRank* method to rank Twitter users to that end [47]. Nevertheless, the behavior of different graph centrality methods and their vulnerabilities when confronted with the Twitter user graph, in general, and Twitter spammers in particular, are still little-known.

Thus, this paper aims to shed some light on this particular issue besides providing some recommendations for future research in the area. As it will be later discussed, user ranking in social networks cannot be an end in itself, but a tool to be used for other tasks. Hence, this author is not considering any *a priori* "good" ranking and, instead, he suggest measuring the performance of the different methods on the basis of two desirable features: on one hand presumed relevant users should rank atop –although the actual ordering among them is irrelevant; and, on the other hand, spammers should achieve lower rankings.

The paper is organized as follows. First of all, a comprehensive literature review is provided. It deals with several rank prestige algorithms (some well-known and others lesser-known) which are applicable to social networks; their known vulnerabilities; and some partially related work and proprietary tools outside the scope of this study. In addition to that, Twitter spam is discussed with a focus on link spam (known as follow spam in Twitter). Then, the different strategies to fight spam in social websites are overviewed. Finally, the research questions are stated and the feasibility of "desensitizing" prestige ranking algorithms against follow spam is analyzed. After that, the experimental framework in which this study was conducted is described: the dataset crawled from Twitter; the elaboration of the subset of relevant and abusive users; and the straightforward nature of the evaluation. Afterwards, results obtained with each of the different ranking methods are discussed along with the implications of the study. Finally, an in-depth analysis of the collected dataset is provided in an appendix: it provides details on the nature of the social network, in addition to some demographical analysis.

[1] Correspondence to: Daniel Gayo-Avello, Department of Computer Science (University of Oviedo) – Edificio de Ciencias, C/Calvo Sotelo s/n 33007 Oviedo (SPAIN), dani@uniovi.es

# 2. Literature review

A social network, despite the current association with online services, is any interconnected system whose connections are a product of social relations or interactions among persons or groups. That way, families, companies, groups of friends, or scientific production are social networks.

Social networks can be mathematically modeled as graphs and, thus, graph theory has become inextricably related to social network analysis with a long history of research. Think, for instance, of bibliometric studies that can be traced back to Lotka [37], Gross and Gross [22], Broadman [7], and Fussler [15], although the work by Garfield [16] is, with no doubt, the one with the highest impact on the daily life of nowadays scholars. However, it is not our aim to provide a survey on this topic; we recommend the reader interested in social network analysis from a Web mining perspective the corresponding chapters from the excellent books by Chakrabarti [9] and Liu [36]. Instead, for the purpose of this paper it should be enough to briefly sketch the concepts of *centrality* and *prestige*.

Both centrality and prestige are commonly employed as proxy measures for the more subtle ones of importance, authority, or relevance. Thus, central actors within a social network are those which are very well connected to other actors and/or relatively close to them; this way, there exist several measures of centrality such as degree, closeness, or betweenness centrality.

While centrality measures can be computed for both undirected and directed graphs, prestige requires distinguishing inbound from outbound connections. Thus, prestige is only applicable to directed graphs which, in turn, are the most common when analyzing social networks.

As with centrality, there are several prestige measures such as indegree (the number of inbound connections, e.g. cites, in-links, or followers), proximity prestige (related to the influence domain of an actor, i.e. the number of nodes directly or indirectly linking to that actor), and rank prestige, where the prestige of a node depends on the respective prestige values of the nodes linking to it –rank prestige is mutually reinforcing and, hence, it requires a series of iterations over the whole network.

Given their importance, and for the sake of clarity, a comparison between the two last prestige measures is provided. Proximity prestige is computed as the mean length of all the shortest paths connecting a given node to the nodes within its influence domain. In other words, proximity prestige measures reach as the mean number of "hops" between a node and all of the nodes linked (direct or indirectly) to it. In contrast, rank prestige takes into account the prestige of nodes linking (direct or indirectly) to a given node –that's why it requires iterative algorithms– and, in some sense, it describes how well connected is a node to other well connected nodes.

Rank prestige is, by far, the most commonly used prestige measure and there exist a number of well-known methods to compute one or another "flavor" of such a measure. In the following subsection we will briefly review the popular *PageRank*, and *HITS* algorithms, in addition to lesser-known (although better targeted at social media) techniques such as *NodeRanking*, *TunkRank*, and *TwitterRank*, besides their weaknesses in different abusive scenarios.

## 2.1. Rank prestige algorithms

### 2.1.1. *PageRank*

*PageRank* [45] is, in all probability, one of the best known rank prestige methods because it underlies the Google search engine [6]. The *PageRank* algorithm aims to determine a numerical value for each document in the Web, such a value would indicate the "relevance" or "authority" of that given document. That value, also known as *PageRank*, spreads from document to document following the hyperlinks –previously it must be divided by the number of outgoing links. That way, heavily linked documents tend to have larger *PageRank* values, and those documents receiving few links from highly relevant documents (i.e. documents with large *PageRank* values) also tend to have large *PageRank* values.

After iterating a finite (in fact a relatively short) number of steps the algorithm converges; at that moment all the nodes within the graph have got a *PageRank* value by means of which they can be ranked. A notable property of the algorithm is that the global amount of *PageRank* within the graph does not change along the iterations but it just spreads from some nodes to other ones. Thus, if the total amount of *PageRank* in the Web was arbitrarily fixed at 1 we could see the *PageRank* value for a given document as a proxy for the probability of reaching that given document by following links at random (that's why *PageRank* is often described as a *random surfer model*). Such a model is described by Equation (1) where $PR(p)$ is the *PageRank* value for webpage $p$, $M(p)$ is the set of webpages linking to $p$ and $L(p)$ is the set of pages linked from $p$.

$$PR(p_i) = \sum_{p_j \in M(p_i)} \frac{PR(p_j)}{|L(p_j)|} \quad (1)$$

Of course, this description is an oversimplification because it makes several unreal assumptions, namely that the Web is a strongly connected graph, and that there are no sink nodes (i.e. nodes with in-links but no out-links). In order to solve this, a modified version including a *damping* or *teleportation* factor is shown in Equation (2): $d$ is the damping probability (usually 0.15), and $N$ is the total number of webpages in the graph.

$$PR(p_i) = (1-d) \sum_{p_j \in M(p_i)} \frac{PR(p_j)}{|L(p_j)|} + \frac{d}{N} \quad (2)$$

### 2.1.2. *HITS*

Hyperlink-Induced Topic Search – *HITS* [30] is another algorithm to estimate the relevance of a document. The method assumes the existence of two different kinds of documents in the Web: *authorities* and *hubs*. An authority is a heavily linked document because each inbound link is a "vote" cast by the user linking that document. Conversely, a hub is a document comprising links to several authorities; therefore, hubs are valuable resources in the Web's ecosystem because they ease users the task of finding relevant information.

Because webpages can exhibit both characteristics every document in the Web has associated two different scores: namely, its authority score and its hub score. It must be noticed that *HITS* is not aimed to be computed across the whole Web graph but, instead, within a query dependent subgraph composed of those documents already satisfying a given query (obtained by means of a standard information retrieval system), plus those documents linked from, or linking to documents in that result set.

Therefore, *HITS* starts with a relatively small Web subgraph and iteratively computes both scores for every page in the graph. As it can be seen in Equations (3) and (4) authority and hub scores mutually reinforce themselves; the authority score for a given page $p$ is the sum of the hub scores of those pages $q$ linking to $p$ ($E$ is the set of edges in the graph) while the hub score for a page $p$ is the sum of the authority weights for those pages $q$ linked from $p$. It must be noticed that with each iteration both scores must be normalized so their squares sum to 1. *HITS*, as *PageRank*, converges after a number of iterations.

Finally, although *HITS* was not devised to compute scores for complete graphs, but rather topic-oriented subgraphs, it can of course be applied to a whole graph and, in fact, that is the way in which we are going to apply it to the Twitter user graph, using the computed authority scores to rank the users.

$$auth(p) = \sum_{q:(q,p) \in E} hub(q) \quad (3)$$

$$hub(p) = \sum_{q:(p,q) \in E} auth(q) \quad (4)$$

### 2.1.3. Abusing *HITS* and *PageRank*

In spite of claims on the original *PageRank* paper about being *"virtually immune to manipulation by commercial interests"*, the fact is that both *PageRank* and *HITS* are prone to manipulation or, at least, they have weaknesses that can be exploited under certain circumstances.

Bharat and Henzinger [5] describe three scenarios were hyperlink analysis methods (a) can be abused, or (b) fail because of wrong assumptions. Such scenarios are: (1) mutually reinforcing relationships between hosts, (2) automatically generated links, and (3) non-relevant nodes.

The first case occurs when a document in a host is linked by many documents from a second host; because each link is counted as a single vote although they are, in all probability, published by the same author, a single individual –the one publishing the links– is earning undue importance. This phenomenon is the underlying base for the so-called *link farms* which plague the Web, and is also somewhat related to Sybil attacks[2] in reputation based systems (as one could consider Twitter).

The second scenario does not describe an abusing situation *per se*, but an assumption made by hyperlink analysis methods that eventually proved wrong: that links are published by human beings where many of them are in fact, automatically generated[3]. Although not totally equivalent, behaviors in social networks such as auto-following users, can for sure bias the results eventually obtained by algorithms such as *HITS* or *PageRank*.

The third and last scenario, namely non-relevant nodes, especially affects *HITS*. Bharat and Henzinger describe how documents not relevant to the query topic can drift the results if they are well connected. In contrast to the previous two scenarios, for which we can find comparable situations within a social network setting, this third one is a little more elusive. In truth, this situation can only be broadly compared with one of the most common spamming behaviors in Twitter, namely getting the more followers the better no matter the relation between the contents promoted by the spammer to the potential interests of the eventual followers.

### 2.1.4. *NodeRanking*

*NodeRanking* [46] can be considered another variation of the random surfer model with authority spreading from one node in the graph to those linked from it. The main differences between *NodeRanking* and *PageRank* are two: (1) it is devised to work on weighted graphs, and (2) the damping/teleportation parameter is not fixed for the whole graph but is computed for each node and depends on the outbound connections of the node (see Equation 5). According to its authors, this feature makes *NodeRanking "able to adapt dynamically to graphs with different topologies."*

Thus, Equations (5), (6), and (7) underly this algorithm. $P_{jump}(p)$ is the probability of damping each node $p$. As it can be seen, nodes with few outbound links have a greater probability of being damped; this could be interpreted as the random surfer getting bored because of the limited set of choices.

$P_{choose}(p)$, is the probability of a page $p$ to be chosen by the random surfer after visiting page $q$ (which, of course, would have a link to $p$). In the original work by Pujol *et al.* this equation employs the edge weight from $q$ to $p$ in the numerator, and the sum of the weights for all of the links departing from $q$ in the denominator. In this work the edges in the social graph are weightless and, thus, we are showing a simplified version of the original $P_{choose}(p)$ equation: i.e. a web surfer visiting a given page $q$ would continue to any of the $p$ pages linked from $q$ with equal probability.

Finally, equation (7) describes how authority is "transferred" from a node $q$ to a node $p$ linked from $q$. As it can be seen, the authority of $q$ is weighted according to the probability of visiting $p$ after $q$ and then it is accumulated on the current authority of $p$. Since this would cause authority values approaching infinity Pujol *et al.* introduced a correcting factor $F_p$ which depends both on the authority of page $p$ and the total authority of the whole graph (this is very similar to the normalization of values required by the *HITS* algorithm).

$$P_{jump}(p) = \frac{1}{1 + |L(p)|} \quad (5)$$

$$P_{choose}(p) = \frac{1}{|L(q)|}, (q,p) \in E \quad (6)$$

$$auth(p) = auth(p) + \frac{P_{choose}(p) \cdot auth(q)}{F_p}, (q,p) \in E \quad (7)$$

[2] A Sybil attack consists of one attacker forging several different identities which are in turn use to promote/link a given resource. The name is after Sybil Dorsett, a woman with dissociative identity disorder and the subsequent book studying her case.

[3] Think, for instance, in the role links such as *Powered by Wordpress*, or *Powered by Apache*, have played in the ranking of their respective websites.

### 2.1.5. *TunkRank*

As we have already noticed, both *PageRank* and *HITS* (in all probability the most commonly applied methods) are prone to manipulation when applied to the Web graph, in general, and to the Twitter user graph (or any other social network graph), in particular. Thus, it could be wise to propose methods tailored to the particular circumstances of social networks. One of such methods is the one originally proposed by Daniel Tunkelang [51] and later named *TunkRank* for obvious reasons.

*TunkRank* defines *influence* as a numerical estimate for the number of people who will eventually read the tweets by a given user (including retweets[4]). The *influence* for a given user is computed from the *influence* of his followers but taking into account two constrains:

First, the *influence* of a user *A* following a user *B* is not transmitted in full to *B*; instead, *A* distributes his *influence* evenly among all of his followees. The intuition behind this is that attention of users is scarce and must be spread; without additional knowledge an even distribution is the most sensible assumption.

Second, a tweet by user *B* won't be read by followers of *A* unless *A* retweets it; therefore, since *influence* is an estimator of the reach of a user's tweets the probability of retweeting ($p$) must be incorporated.

Tunkelang suggests computing users' influence recursively and argues that, although infinite for graphs containing cycles, it would converge as powers of $p$ approach zero. In fact, shortly after describing this method an implementation for *TunkRank* was publicly released[5].

Equation 8 shows the way in which *influence* for a given user $X$ can be computed. As it can be seen, the equation incorporates all of the aforementioned constrains. The even distribution of attention to followees is expressed by means of the denominator, where $|Following(Y)|$ stands for the number of users is following user $Y$. The probability of retweeting, $p$, appears in the numerator; again, a simplifying assumption was made: such a probability is equal for all of the users in Twitter.

Up to now no rigorous analysis of *TunkRank* has been performed; however, it seems plausible that, given its remarkable similarity to *PageRank*, it would suffer from many of the weaknesses described above (e.g. Sybil attacks, auto-following, and link spamming in general). Hence, to the best of our knowledge, this is the first thorough scholar analysis on *TunkRank*.

$$Influence(X) = \sum_{Y \in Followers(X)} \frac{1 + p \cdot Influence(Y)}{|Following(Y)|} \quad (8)$$

### 2.1.6. *TwitterRank*

*TwitterRank* [52] is an extension to the *PageRank* method which, in addition to link structure, takes into account the topical similarity between users in order to compute the influence one users wield onto the others. In that sense, *TwitterRank* is a topic-sensitive method which ranks users separately for different topics. Thus, in order to rank users globally (i.e. with topic independence) one should aggregate every *TwitterRank* value weighted according the difference topic importance within the corpus.

It must be noticed that, in addition to this, the transition probability among connected users heavily relies in both the topical similarity between users, and the number of tweets published not only by the followee, but by all the followees the follower is connected to. Certainly, these features make of *TwitterRank* a highly flexible method which, in theory, could easily follow topic drifts. However, we feel that such a degree of flexibility makes the algorithm difficult to scale to the number of users and tweets that are published on a daily basis[6].

---

[4] A retweet is a tweet from a given user repeated by other user verbatim and including attribution to the original author. For instance, let us suppose the user `alice` posts a tweet with the text "`Hello world!`" and that the user `bob` wants to retweet it to his followers; he would just have to post "`RT @alice: Hello world!`" In that tweet, there `RT` stands for retweet while the `@` followed by the user name is what in Twitter parlance is called a mention. Mentions are messages publicly addressed to a given users (in contrast to private messages which are called "direct messages" in Twitter).

[5] http://tunkrank.com

Because of this, and for the sake of better comparison with the rest of rank prestige, we employed a slightly modified version of *TwitterRank*. The differences are the following ones: (1) instead of computing a different *TwitterRank* value for each user and topic to be later aggregated across topics, we aimed to compute just one *TwitterRank* value without relying on any topic. (2) We also changed the topical similarity measure to compare users. Instead of applying Latent Dirichlet Allocation (LDA) to find the topics, then obtain each user's distribution, and finally compute Jensen-Shannon Divergence between users' distributions, we decided to apply the much more usual cosine similarity. And lastly, (3) we simplified the way to compute the *damping/teleportation* parameter. In the original paper it was computed from the matrix of users and topics obtained by means of LDA; we, in contrast, use the ratio between the number of tweets published by a given user and the total number of tweets in the corpus.

Equations (9) and (10) provide a description of our implementation of *TwitterRank*. $TR(u)$ is the *TwitterRank* value for user $u$; $\gamma$ is the probability of teleportation, a constant value between 0 and 1 for the whole graph –we used the commonly applied value of 0.15; $P(u_j,u_i)$ is the transition probability from user $u_j$ to user $u_i$; $|\tau_i|$ is the number of tweets published by user $u_i$, and $|\tau|$ is the total number of tweets published by all the users. Lastly, $sim(u_j,u_i)$ is the similarity between users $u_i$ and $u_j$ which, as we have already said, was implemented as cosine similarity.

$$TR(u_i) = (1-\gamma) \sum_{u_j \in followers(u_i)} P(u_j,u_i) \cdot TR(u_j) + \gamma \cdot \frac{|\tau_i|}{|\tau|} \quad (9)$$

$$P(u_j,u_i) = \frac{|\tau_i|}{\sum_{a:u_j follows u_a} |\tau_a|} sim(u_j,u_i) \quad (10)$$

Hence, *TwitterRank* is indeed an extension of *PageRank* which takes into account the topical similarity between users to weight the transitions among connected users, in addition to the number of tweets the different followees publish to establish the influence a user has on its followers.

## 2.2. Other methods and tools to compute "influence"

Apart from the previously described algorithms there have been many other approaches to inferring influence in so-called Web 2.0 environments. Most of such approaches rely not only in the user graph, but they also require additional information such as user actions (e.g. joining a group, uploading a picture, tagging a resource), or the resources and tags collected and labeled by the users in the network.

For the interested reader we recommend the works by Noll *et al.* [42] and Goyal *et al.* [20]. The first one describes the SPEAR algorithm (SPamming-resistant Expertise Analysis and Ranking) which processes data from a collaborative tagging system (e.g. del.icio.us or bibsonomy) to find the most valuable resources and users by means of a mutually reinforcement method. The second one describes different models to determine influence among users by exploiting both the social graph and the actions performed by the users within the service.

With regards to micro-blogging services like Twitter, there are a number of interesting proposals to find authoritative users exploiting idiosyncratic features of the service such as retweets or mentions.

In a seminal work, Cha *et al.* [8] compared a simple graph-derived measure of influence (indegree) with two others better adapted for Twitter: retweets and mentions. According to those authors, users with high indegree (i.e. large numbers of followers) do not necessarily produce large numbers of retweets or mentions; in other words, indegree does not warrant message spreading. In addition to that, Cha *et al.* found that users with large numbers of followees or with a high tweeting rate were usually robots and spammers.

---

[6] The paper originally describing *TwitterRank* employed a dataset consisting of one million tweets published by 6,748 users mainly located in Singapore. The number of users the original authors had to rank is several orders of magnitude below the size of the dataset to be employed in this study (1.8 million). In addition to that, given the homogeneity of the user group and the size of the tweet corpus, it seems pretty clear that the number of possible topics would also be well under the number of topics that could arise in the dataset to be used in this study (27.9 million).

Bakshy *et al.* [2] followed a different approach to estimate how influential users in Twitter are: they tracked the diffusion of URLs across communities of users, and studied the relation between the length of each diffusion cascade and a number of features of the user initiating it. The features were the number of followers and followees, the number of tweets, the account's date of creation, and the past influence of the user which was based on the length of the cascades started by that user in the past. By training on historical data Bakshy *et al.* tried to predict the length a cascade initiated by a given user would reach. Although they found that past influence and large follower base are necessary to start large cascades they are insufficient and, according to them, predictions of cascade lengths were relatively unreliable.

Romero *et al.* [47] proposed the *Influence-Passivity (I-P)* algorithm which is closely related to *PageRank*, *HITS*, or *TunkRank*. The *I-P* algorithm weights the edges of the social graph according to user interactions, concretely, retweets. The underlying intuition is very appealing: users' influence depends on the passivity of their followers and, conversely, users' passivity depends on the influence of their followees. For each pair of users, acceptance and rejection rates are computed for the follower; the former is the ratio of received messages s/he retweets while the later is the amount of influence s/he rejects. This way, the passivity of a user depends both on his rejection rate and the influence of his followees, while his influence depends on the acceptance rate and the passivity of his followers. As in *HITS*, both scores are computed iteratively.

To test their algorithm, Romero *et al.* compared the correlation between influence scores (as computed with the *I-P* algorithm) with clickthrough data on URLs promoted in Twitter (i.e. they used clicks as a proxy measure for attention). According to them, the *I-P* algorithm outperforms *PageRank*, H-index, follower count, and number of retweets.

However, Gayo-Avello *et al.* [17] argued later that Romero *et al.* missed the confounding effect of the number of followers in both the influence and the clickthrough data. When correcting the data for audience size, Gayo-Avello *et al.* did not found any significant correlation between influence computed by means of the *I-P* algorithm and clicks on URLs. However, they did found a significant correlation between clicks and *PageRank* and *TunkRank* scores. In addition to that, *TunkRank* outperformed *PageRank* and a new algorithm devised by Gayo-Avello *et al.* outperformed both of them. That new method does not employ graph data but just relies on the mentions received by the users and their follower counts to compute a dynamic measure of influence which positively (and significantly) correlates with clicktrough data in promoted URLs.

Finally, Kwak *et al.* [32] describe a method to discover influential users in Twitter by analyzing the way in which information is diffused across the network. That is, they do not only consider if a user is following another one, but which new pieces of information s/he discovers via that followee, and how that user propagates (or not) that new information.

Finally, and for the sake of completeness, it must be noted that there also exist some proprietary tools which claim to determine different kinds of influence in Twitter. Among them we can mention *Tweet Grader* (`http://tweet.grader.com`), *Klout* (`http://klout.com`), *PeerIndex* (`http://www.peerindex.com`), or *Twitalyzer* (`http://twitalyzer.com`). Each of them provides one or more metrics which, purportedly, portray the influence or authority a user has on other Twitter users. All of them claim to take into account signals such as the number of followers, the follower-followee ratio, the number of retweets and mentions received by the user, or the connections with other users (and their respective scores). Needless to say, details about how each of these tools computes its metrics are undisclosed and, therefore, they cannot be neither replicated nor thoroughly analyzed.

None of the aforementioned methods was considered for this study. The techniques devised in [20][18] or [42] could (and should) be adapted to microblogging services such as Twitter but such a goal is out of the scope of this paper. In [2] and [32], the social graph is not the most important piece of data, the way in which information flows through it is, instead, key. With regards to [17] and [47] the behavior of the users (either retweets or mentions) is as important as the social graph or, as is the case of [17], it simply replaces it. Finally, [8] just made a point against the use of simplistic graph measures (indegree) as proxies for influence but did not make any attempt to correlate retweets and mentions with other external measures of influence.

Thus, for the purpose of this study, we decided to focus on those methods just relying in the social graph, namely *HITS*, *PageRank*, *NodeRanking*, *TunkRank*, and *TwitterRank*, in addition to a new method proposed by the author of this paper and that is introduced in the following section.

## 2.3. Twitter spam

*"Spam is bad. The amazing degree of unanimity that greets such a simple declaration is, paradoxically, the biggest impediment to progress in anti-spam standards."*

(N. Borenstein, co-creator of MIME)

The reader has dealt, in all probability, with a fair share of unsolicited e-mails on a great variety of topics, found in search results weird webpages that no human could have possibly written, and perhaps read amusing (and totally off-topic) comments to blog posts. All of those experiences are qualified under the broad term "spam".

Twitter is no exception, and spammers have been targeting this service for a number of years. Of course, Twitter's terms of use forbid spamming and there are a number of behaviors the company considers as such: "*Posting harmful links to phishing or malware sites, repeatedly posting duplicate tweets, and aggressively following and un-following accounts to attract attention"* [11]. Twitter provides its users with a number of tools to pointing out spam accounts which are later checked for misuse and eventually suspended.

Needless to say, spammers persist and, as in other online services, the adversarial nature of their behavior makes spamming a popular research topic. Indeed, several papers on this regard have been recently published, such as [4], [28], [34], [48] or [55] just to cite the most relevant.

For instance, Benevenuto *et al.* [3] considered spammers those users tweeting URLs unrelated to the topic of the tweet. Interestingly, they also state the following: *"we do not consider content received through the social links as spam since users are free to follow the users they want."* Therefore, only tweets addressed to a user not following the spammer would be considered by those authors.

Lee, Caverlee and Webb [34] described a number of "families": from the purportedly self-describing "pornographic spammers" and "phishers" to "promoters" (users tweeting a mixture of "legitimate" tweets and tweets about online marketing) and "friend infiltrators" (users who massively follow other users to reach a certain amount of followers and only then spread spam).

This is not, however, the only classification for Twitter spammers. Stringhini, Kruegel and Vigna [48] distinguished four categories of spammers in social networks (including Twitter): "displayers" (users not posting spam but displaying spam on their profile), "braggers" (users posting spam), and "posters" and "whisperers" (the former are users addressing spam to other users in plain sight while the later use direct –and private– messages).

To the aforementioned categories Irani *et al.* [28] added "trend-stuffers": spammers posting tweets including hashtags corresponding to the current trending topics. Such a behavior allows spammers to reach huge audiences because trending topics appear prominently in Twitter's interface and by clicking on them users can read recent tweets on that particular topic.

Needless to say, there have been a number of attempts to characterize and eventually detect spammers in Twitter. For instance, Leavitt *et al.* [33] suggested the follower to followee ratio as a measure to spot "stereotypical spammers". According to them, if a user's ratio approaches zero (i.e. relatively few followers versus relatively many followees) would imply that user is a spammer. Unfortunately, spammers are far from stereotypical and such a method simply does not work: according to both [4] and [55] spammers have rather high numbers of followers.

Hence, machine learning is the most popular approach: it requires a labeled set of legitimate users and spammers to then train a classifier on a number of features obtained from the users in the sample. Among the aforementioned researchers [4], [28], [34] and [48] followed that approach using different machine learning methods and achieving different precision values: 70% [4], 82% [34] or 97.5% [48].

It is worth noting that Yardi *et al.* [55] showed that a rather simple algorithm based on the presence of URLs, spam keywords and certain patterns in the user names, was able to achieve 91% precision [55]; this is remarkable because such an algorithm makes unnecessary preparing a sample of spammers to train the classifier.

We will close this section with an operative definition for Twitter spammers consistent with all of previous works albeit necessarily broad:

- A Twitter spammer is a purported user –actually a bot or an instrumented account– with an underlying profit-making goal which requires persuading other users to visit one or more websites outside Twitter.
- To achieve that goal spammers' tweets tend to contain URLs (usually in a higher proportion than the rest of the users).
- Depending on the profit-making method, the message can range from transparent to overtly fraudulent. In other words, if the promoted website is not totally illegal the tweet text can be related to the contents of the website (e.g. weight-loss and diet websites), while if the website is malicious the tweet text would be unrelated to the contents of the website (e.g. phishing websites).

- To maximize the probability of users visiting the promoted websites, spammers require both frequent repetition of their messages and huge audiences.
- To achieve the former they tend to rely on minor modifications to the message; this can be achieved by changing the URL using URL shorteners and/or addressing each tweet to a different user.
- To achieve the later there are two main approaches: (1) spammers can aggressively follow hundreds or thousands of users assuming some of them will follow-back thus becoming followers. (2) Spammer can include in their tweets hashtags corresponding to current trending topics or habitual Twitter memes. Given that users tend to search Twitter by using such hashtags as queries, spammers' tweets would appear as search results having, thus, an opportunity for the promoted website to get clicked.
- Those spammers engaged in the first approach to get huge audiences risk showing a telltale sign: an oddly low follower to followee ratio. Because of this, it's not uncommon that spammers do not only aggressively follow other users but they also unfollow them shortly afterwards.

Previous description satisfies all of the assumptions made in the aforementioned works about spammers and, still, it shows that there are many different kinds of Twitter spammers. They differ on the way they make profit (legally or illegally), the nature of the product or service they promote, the way in which they reach their audiences, or their method to deliver their promotional messages.

# 3. Research motivation

## 3.1. Spam countermeasures

From the literature review regarding Twitter spam it can be concluded that false negatives (uncatched spammers) and false positives (legitimate users labeled as spammers) are unavoidable and, even worse, as it was stated in [3], [11] or [55] the "war on spam" is a never ending fight because every deployed anti-spam measure will be eventually overcome by spammers.

However, all of those works approach Twitter spam as a detection task while there are other approaches to fight it without requiring a detection phase. Here, we will refer to the excellent survey by Heymann, Koutrika and Garcia-Molina [26]. According to them, there are three possible countermeasures to spammers: detection, demotion and prevention.

As it has been shown, detection methods have two main problems: false positives and negatives, and knowing for sure that spammers will evolve to pass the detection filters.

Prevention methods aim to make spam posting more difficult. According to Heymann *et al.* this would require limiting automated interaction. CAPTCHAs are one of such measures and they mention it as a way to avoid automated account creation; nevertheless, it would be rather annoying requiring CAPTCHAs for every interaction in Twitter. In addition to that, although prevention methods can be an interesting approach to limit the occurrence of spam in Twitter in the future, they would require major changes to Twitter APIs and user interface and, thus, they seem to be of no immediate use in the short term.

Finally, demotion methods *"attempt to lower the ranking of spam in ordered lists"* in the words of Heymann *et al.* That is, the system does not prevent nor detect and remove spam or spammers; instead, spam (or spammers) should have their prominence reduced. In this sense, they specifically discussed Web search engines which suffer the so-called "spamdexing". That kind of spam is defined by Gyöngy and Garcia-Molina [23] as *"any deliberate human action that is meant to trigger an unjustifiably favorable relevance or importance for some web page, considering the page's true value."*

A number of similarities can be drawn between Web search and Twitter search scenarios. The volume of published information is huge and not curated; anyway, worthwhile information could be available provided it could be adequately ranked; hence, there exist an incentive for cheating the eventual ranking algorithm and, therefore, spamdexing is a plausible threat.

Under such circumstances demotion-based methods are much more convenient than detection or prevention based ones for a number of reasons:

- Prevention methods consist of putting barriers to publishing which, at a minimum, are bothersome for legitimate users.

- Detection methods are enforced only once spammers have plagued the community and besides they eventually turn into an "arm race" which makes spammers better and better fitted and, hence, difficult to detect in the future.
- Demotion based methods are akin to "security by design" approaches to software building. That is, spammers are taken for granted and the ranking algorithms are envisioned to minimize both the impact of spamming actions and collateral effects in legitimate users.

Needless to say, up to now, ranking algorithms have not been devised to be "spam proof by design"; consequently, and given that Twitter search is still a rather unexplored field[7], it is pertinent to study the way in which an algorithm could rank relevant users in Twitter while at the same time demoting spammers.

## 3.2. The importance of reciprocal linking in Twitter spam

An operative definition of Twitter spammers was provided above. There, a number of features of spammers were described but there is one that deserves closer attention because (1) it is instrumental to the way in which most spammers operate in Twitter, and (2) it is somewhat related to the problem of link farming in the Web.

As it was said, spammers require huge audiences to monetize their messages. Some spammers target so-called *trending topics* in order to "inject" their tweets in the search results of users looking for information about such topics. However, the most popular method and, by far, the most widely reported in the literature [3][12][13][18][19][21][40][53][54] is that of following users expecting to be followed back.

This is also consistent with the findings of this author: Table 4 reveals that spammers use of hashtags is significantly higher than that of other users; however, spam tweets with hashtags are far from being the majority of their tweets and, thus, most of the spammers are counting on his followers as their main audience.

This behavior has been denoted by Twitter as "follow spam" [53] and it exploits the fact that many users tend to follow back any other user who starts following them [3][40].

Indeed, it has been reported that sophisticated spammers do not only target those users more prone to follow back [18][19] but they also unfollow those that do not reciprocate the following [3][12][40].

Hence, although reciprocation could be thought of as a sign of mutual interests or even actual friendship, it has been actually found that spammers and other fraudulent accounts exhibit higher reciprocation than average users [18][54].

In fact, spammers are getting their followers from a relatively small subset of Twitter users (about 100,000 accounts) who have huge follower bases and routinely follow back all of them [19]. In this regard, follow spam is related to link spam in the Web (a subclass of spamdexing, see [23]) and could be tackled with in similar ways: indeed, Ghosh et al. [19] described *CollusionRank*, an algorithm similar to *TrustRank* [24], to assign spamminess scores to Twitter users starting from a seed of known spammers.

So, in short, among the different features that can help to characterize spammers in Twitter, reciprocal linking, or follow spam, seems to be instrumental and, as it has been said, algorithms based on the fact that spammers rely on such reciprocal links have successfully spotted unknown spammers [19].

Needless to say, such linking behavior can also be seen in other users and, hence, its mere presence would not immediately point out a user as a spammer. Moreover, as it has been argued before, although spam detection is unavoidable in the Web, Twitter is relatively new and there is still room for taking a demotion based approach against spam which, as discussed above, is far more convenient than a separated detection phase.

## 3.3. Research questions

Social networks are increasingly gaining importance in the day-to-day living of Internet users, and the contents they provide can be exploited to provide up-to-date information (the so-called real-time web). Needless to say, because of the ease of publishing any content, anytime by anyone, it is ever more important to have a way to separate trustworthy/relevant/authoritative sources of information from the untrustworthy/irrelevant/un-authoritative.

---

[7] As a matter of fact, the Text REtrieval Conference (TREC) introduced a track on IR in microblogs for the first time in 2011.

Given the prior success of applying rank prestige algorithms to bring order to the Web, it seems appealing to do the same with the user graphs from social networks.

However, as it has been shown above, ranking algorithms can be abused (for instance through spamdexing and link farming) and, in fact, spammers following similar approaches (the so-called follow spam) are already operating in Twitter.

Besides, although there is a large body of work regarding spam detection in the Web, spam demotion can be a more convenient approach given that spammers have not been abusing Twitter for too long.

In this sense, there is still a "window of opportunity" to study to what degree ranking algorithms proposed up to date are spam proof (or at least spam tolerant), and the feasibility of trying to make them more robust to spammers.

Also, and as it has been aforementioned, reciprocal linking (follow spam in Twitter, link spam in the Web) is instrumental for spammers and, hence, that behavior (typical in spammers but also in other users) has been put at the core of the research described in this study.

Therefore, the main research questions addressed in this study can be stated as follows:

1) How vulnerable to follow spam are common graph methods when applied to user graphs from online social networks?
2) Is it possible to "desensitize" such algorithms in a way that avoids to detect and filter out spammers after computing the ranks, but that, instead, take into account their presence and minimize their influence?

It must be said that in addition to the graph centrality methods studied in the Literature Review this report includes an analysis of a variation of the popular *PageRank* method which, purportedly, should be less sensitive to link abusing in social networks.

Such a variation, which will be thoroughly described in the following subsection, relies on a deweighting factor computed from the reciprocal links between users. In contrast to previous solutions, this variation incorporates information about reciprocal links during the ranking computation to avoid a prior spam detection phase while, hypothetically, demoting spammers anyway.

## 3.4. A rationale for "desensitizing" prestige ranking methods against link spamming

As we have previously exposed, one of the simplest prestige measures in a network is indegree which translated to Twitter terms is the total number of people following a user. A priori it seems a reasonable approach: as Leavitt *et al.* pointed out, *"the more followers a user possess, the more impact he appears to make in the Twitter environment, because he seems more popular."*

Users such as Oprah Winfrey (3.1M followers), CNN Breaking News (2.9M followers), or TIME (2M followers)[8] are almost expected to have such huge number of followers; after all, they are opinion-makers and mass media. One could even find reasons to explain the number of followers for Ashton Kutcher (4.5M), Britney Spears (4.4M), or Lady Gaga (2.8M): they are celebrities, fans are eager to know about their idols and feel they are in contact with them, etc. Which is harder to understand is how can spammers have far more followers than average users (Yardi *et al.* 2010; see also Section 4.2 for more details on this).

An explanation for such a phenomenon has been provided above: As with any other social environment, Twitter has seen the emergence of its own etiquette and, for many users, following back a new follower is considered "good manners". Of course, such a behavior is not a problem *per se* and, in fact, it makes perfect sense: if somebody starts following you, it means (in theory) s/he is interested in what you are tweeting about; probably both of you have some common interests and, thus, it would be a good idea to follow-back that user to see what s/he is publishing.

Needless to say, many users are just following back their new followers as a matter of custom and many of them are using different tools and scripts to auto-follow back their followers[9]. Once spammers took notice of this behavior it was relatively easy to get followers: spammers just needed to massively follow other users. It must be said that Twitter consider this a violation of their terms of use but spammers (and many users in general) are using this and other related methods to increase their follower count.

[8] All the follower counts in this section are as of mid February 2010.

[9] Twitter had once got an auto-follow feature that was available under request. It was later disabled because of its harmful potential.

Hence, it seems that number of followers is not to be trusted when trying to infer a user's relevance. Indeed, we have already said that Leavitt *et al.* [33] suggested that the ratio between the number of a user's followers and the number of his followees is a better metric.

According to these authors, if this ratio approaches infinity the user is a successful broadcaster, *"moving content to other users in the environment."* For instance, Oprah has a follower-followee ratio of $1.67 \cdot 10^5$, CNN Breaking News of $1.04 \cdot 10^5$, TIME $2.19 \cdot 10^4$, Ashton Kutcher $1.47 \cdot 10^4$, Lady Gaga 18.08, and Britney Spears 10.43.

To the contrary, if the ratio tends to zero Leavitt *et al.* qualify that user as a *"stereotypical spammer"*; nevertheless, [4] and [55] have found that spammers do not fulfill such a stereotype. In fact, the follower-followee ratio for spammers tends to be close to 1 and, actually, many of them manage to have a positive ratio indeed. With regards to common users it varies widely but, as an example, the ratio for this author is a meager 1.30.

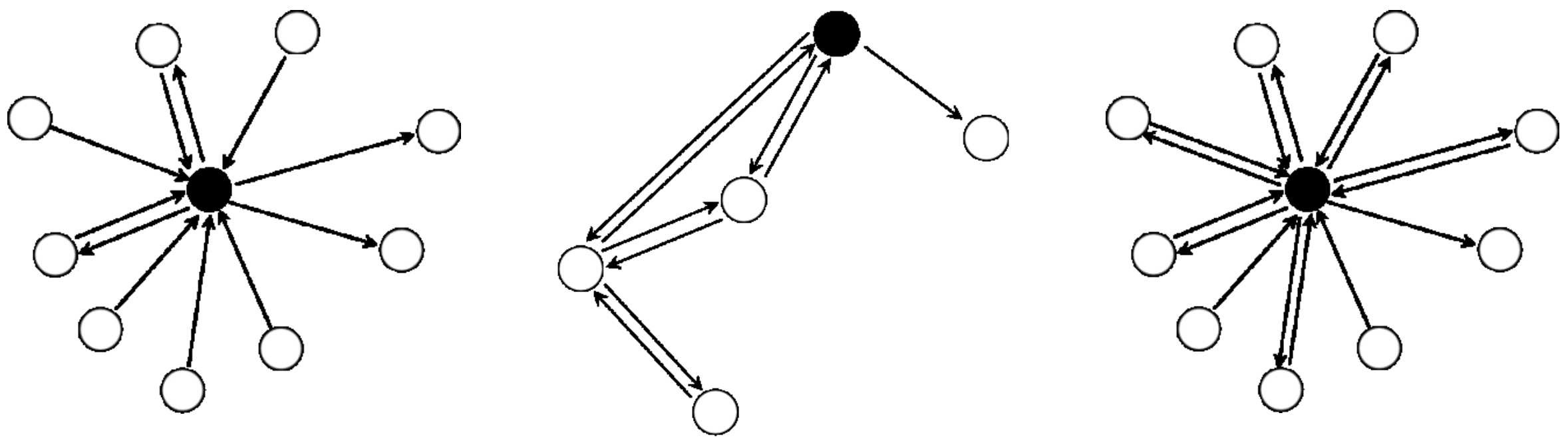

**Figure 1.** Three archetypical scenarios in a social network like Twitter. From left to right: an opinion-maker, a user with a close group of friends, and a user making connections by means of link exchanges (i.e. a probable spammer).

Figure 1 illustrates these three common scenarios. The first one is the typical case for a celebrity/mass-media: s/he has got lots of followers plus a few followees[10]. The second case shows the archetypical user with a close group of friends or relatives in addition to a few followees outside that group. In this scenario there are lots of connections between the users in the close group and relatively few outside. Lastly, the third scenario shows a user who has managed to get followers by mass following other users and who, in fact, has managed to have a few more followers than followees[11]. In these toy examples the follower-followee ratio is 1.75 for the heavy-followed user, 0.67 for the user with a close group of friends, and 1.14 for the presumptive spammer.

Nevertheless, let's put aside spammers for a moment so we can pay a little more attention to the follower-followee ratio. Is it just an ad hoc heuristic? Or, on the contrary, does it provide any sensible (and useful) reading? In our opinion it can be interpreted as the user's "value" regarding the introduction of new original information from the outside world into the Twitter global ecosystem. Those users which publish valuable tweets get followers who do not mind if those users are "impolite" (i.e. they do not follow back); that way they have huge number of followers but small numbers of followees and, thus, their ratios tend to be large. On the other hand, users who tend to discuss relatively personal matters with their close group of acquaintances do not get large audiences and, in turn, their ratios tend to be small (even close to zero if they follow lots of people).

But there is the spam problem. How should we tackle with it? As it was pointed above, the answer may lie on reciprocal connections, those where two users are following each other. As we have said, many users consider this a sign of politeness but, many, especially those with huge numbers of followers, simply cannot follow-back everybody (not if they want to actually read what their followees are writing). Spammers, however, are no reading tweets and, thus, they have no constrain in the number of people to follow; especially if they aim to get a new follower in reciprocity.

---

[10] Just because you are Oprah or Ashton it doesn't mean you don't need to follow CNN.

[11] Probably by unfollowing users after s/he has got a follow-back link after following those users in first place.

In other words, reciprocal links should be under suspicion because many times they are used as "counterfeit currency" to increase the followers count. In consequence, we propose the *follower-followee ratio with discounted reciprocity* –see Equation (11)– which, in our opinion, captures many of the subtleties of linking in social networks.

$$ratio_discounted = \frac{followers - reciprocal}{followees - reciprocal} \quad (11)$$

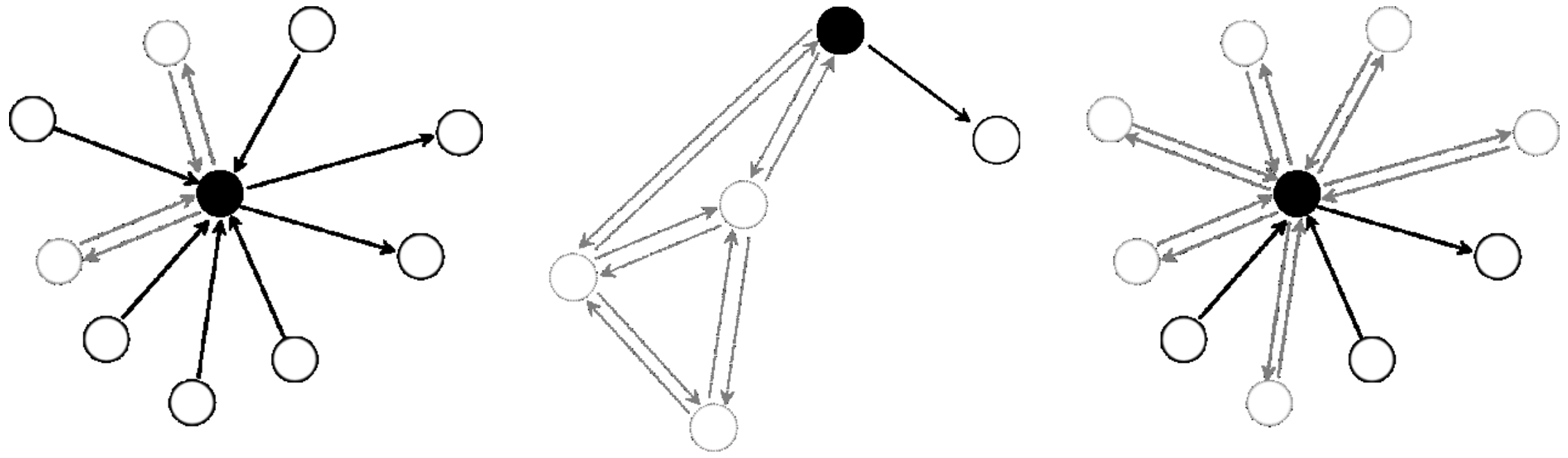

**Figure 2.** The same three scenarios from Figure 1 but "discounting" reciprocal connections. As it can be seen heavy-followed users are hardly affected while those users with many reciprocal connections such as small groups and spammers tend to "lose" most of their inbound links.

Figure 2 shows the same three scenarios from Figure 1 but reciprocal connections are shown in a lighter shade. Those users with many of such connections, namely spammers and common users, tend to "lose" most of their followers, while heavy-followed users are virtually unaffected. The crux of the matter is that, although for users in small close communities this could go undetected, spammers would find such a thing undesirable.

On the other hand, putting under suspicion all reciprocal links seems a bit obnoxious; that's why we don't suggest using this ratio "as is" but, instead, employing either the follower-followee ratio or the discounted version depending on the possible outcome: if a user would "benefit" of using the original ratio then we use the discounted one, and vice versa. Because of that, the whole name for the proposed ratio is in fact *followers to followee ratio with paradoxical discounted reciprocity* (Equation 12). For sure this can look a bit nonsensical at first but, as we are about to show, it makes perfect sense.

First, let's take two users: one of them, *legit*, has 34,000 followers and 300 followees with 200 reciprocal connections; the other one, *spammer*, has 25,000 followers and 30,000 followees with 20,000 reciprocal links. The ratios for them would be 113.33 and 0.83, respectively. The ratios with discounted reciprocity would be 338 and 0.5, respectively. If the users could choose which ratio they prefer to describe themselves it seems clear that *legit* would prefer discounted reciprocity which is larger, while *spammer* would prefer the raw ratio.

However, such selfish decisions are contrary to the interpretation of the ratios. That is, *legit* would prefer assuming his reciprocal connections are not legitimate but follow spam. On the other hand, *spammer* would prefer to count all his reciprocal connections as truly legitimate while, in all probability, this is not the case (you cannot simply follow thirty thousand people). However, both users could get us to apply the opposite ratio. For instance, *spammer* could unfollow 20 thousand people to achieve a 2.5 ratio; however, such massive unfollowing would probably make him lost most of his followers. And what about *legit*? He could massively follow most of his followers but, independently of the ratio to apply, such an action would only reduce his final value. Thus it does not seem that neither *legit* nor *spammer* would change their connections just to force a different way to compute their respective ratios.

$$paradoxical_discounted(p) = \begin{cases} \dfrac{followers(p)}{followees(p)} & if\ followers(p) > followees(p) \\[2ex] \dfrac{followers(p) - reciprocal(p)}{followees(p) - reciprocal(p)} & otherwise \end{cases} \quad (12)$$

Finally, the paradoxical discounted ratio is not aimed to be directly applied to users in the graph but, instead, as a weight within another algorithm such as *PageRank*. For the sake of illustration, we will discuss here how to modify the original *PageRank* equation (see Eq. 1) in addition to other minor considerations to take into account for implementation.

Equation 13 corresponds to the incorporation of the paradoxical discounted ratio to *PageRank*. As it can be seen, for each page $p_j$ linking to page $p_i$ its *PageRank* score –$PR(p_j)$– is weighted according to the number of outgoing links from $p_j$ and, in this modified version, its paradoxical discounted ratio. Please note that the value for the ratio is normalized according to the maximum reached by all the nodes in the graph (remember that the value is always positive but unbounded).

By doing this, the ratio is playing the role of an externality, that is, it has an impact not on the agent responsible of its value but on third persons. In other words, a user is not affected by his own ratio, his followees are (the prestige they received is de-weighted). From an economic point of view this is an appealing feature because individual spammers do not have any incentive to modify their personal behavior, although as a group –they tend to weave tight networks– all of them lose out.

The only final consideration when incorporating the paradoxical discounted ratio to a given algorithm is that the scores for each page must be normalized after each iteration (otherwise they would decrease towards zero).

$$PR(p_i) = \sum_{p_j \in M(p_i)} \frac{PR(p_j)}{|L(p_j)|} \cdot \frac{paradoxical_discounted(p_j)}{max_paradoxical_discounted} \quad (13)$$

# 4. Research design

The main goal of this study was to compare the performance of different rank prestige algorithms when applied to social networks. To attain that, a relative large dataset was needed, in addition to an objective criterion against which to compare the performance of the different methods. The dataset, described in the following subsection, was crawled by the author from Twitter. Within that dataset two different subsets were prepared: one of presumed relevant users and another of abusive users.

The second subset was needed because the research questions underpinning this study deal with the vulnerability of centrality methods to spammers and the feasibility of “desensitizing” them to their abusive behavior. As it was said a desirable feature of any ranking algorithm would be that of demoting abusive users and, the lower the scores for spammers the better the algorithm.

Nevertheless, although demoting spammers is a desirable feature for a ranking algorithm it is not its main goal; ranking relevant users atop is: the higher the ranks for those users the better the algorithm. That is the reason for the first subset of users which, along with the subset of abusive users, will be described below.

Thus, this section describes the dataset employed for the research, the way in which subsets of presumably relevant and abusive users were extracted, and the simple evaluation method applied to compare the different algorithms.

## 4.1. Dataset description

We relied on the Twitter search API to create the dataset used in this study. To that end, we employed a query composed of frequent English stop words (e.g. *the*, *of*, *and*) in addition to forcing the results to just include tweets written in English. That query was submitted once every minute from January 26, 2009 to August 31, 2009. Needless to say, the crawling was not flawless and, in fact, there were 18 days with minor network blackouts and 3 days with no tweets at all (April 26, and August 22 and 24). All in all, we collected 27.9 million English tweets corresponding to 214 days.

As we have already implied, users in Twitter can involve themselves in relationships with other users. Thus, a user can *follow* another one so that the first user can receive the tweets published by the second. This way, users can have, in Twitter parlance, *followers* and *friends*. Please notice that friends are no more than the persons a given user is following; hence we are using the term *followee* instead [27].

Using that information on followers and followees, Twitter can be represented as a directed graph. In order to build such a graph we tried to obtain followers and followees for each of the 4.98 million users appearing in the dataset. To do that, we employed the so-called *Social Graph Methods* provided by the Twitter API. This second crawl was performed after the first dataset was collected and took several months.

For the final graph we did not take into consideration links from/to users not appearing in the sample and we also dropped isolated users. In addition to this, a substantial amount of user accounts were suspended[12] at the moment of the second crawl and, hence, no information on these users' followers and followees was available. Lastly, because of the unavoidable network problems, coupled with the fact that we pushed the API a little too far, the information for a noticeable amount of users was not eventually crawled.

The finally collected Twitter graph consisted of 1.8 million users; that is, 36% of the users within the original sample. It is bigger than other datasets reported in the literature (e.g. Java *et al.* [29] –about 90,000 users– or Choudhury *et al.* [10] –200,000 users) but smaller than others (e.g. Kwak *et al.* [31] –41.7 million users, the whole Twitter graph as of July 2009). Anyway, we think it is a fairly substantial sample given that, at the moment of collecting the dataset, the number of U.S. Twitter users was estimated between 14 [43] and 18 millions [44] and, thus, our sample is 10% that size and covers a little more than 4% of the whole Twitter graph. An in-depth analysis of this dataset is provided as an Appendix for the interested reader.

## 4.2. Data preparation: relevant users

The goal of applying a ranking algorithm to a social network is to find the most relevant users that should appear atop. Hence, if only we were able to find a ranked list of users to be used as a ground truth, it would be straightforward to compare the different algorithms. Needless to say, we could rely on lists similar to those of Forbes magazine but there are a number of problems with them:

- Although there is some overlap between such lists and the Twitter user base, not every person in those lists is using Twitter and not every relevant Twitter account is a person.
- Despite the claims made by their proponents about using a methodology they are highly debatable at least and they would be hard to justify in a scientific paper.
- The very idea of ranking users may be considered preposterous even when focusing in a very specific field. For instance, what would be a "right" ranking for actors and actresses? Or for entrepreneurs?

That does not mean, however, that there is no way to find a subset of users against which to test the performance of the different algorithms. It means that, firstly, we should focus on Twitter's ecosystem and, secondly, we do not need to have that group internally ordered or ranked but, instead, we need to have that group ranked atop as a whole to say that a ranking algorithm is properly working.

From such a perspective there is a group of users that can help our purposes: celebrities. Such a group is composed of actors, news anchors, politicians, sport figures, etc. and, depending on the different interests of the audience, they can be considered relevant. Moreover, all of them share a problem: to warrant their followers that they are who they seem. Besides, the problem is not only theirs but also of Twitter: by not properly handling deceptive accounts of celebrities the company could very well get involved in libel lawsuits.

Hence, Twitter is proactively verifying accounts to check they actually belong to the celebrity they claim. Verified accounts display a characteristic "badge" in Twitter's interface and can be also checked thanks to the `@verified` Twitter account.

Needless to say, the number of verified accounts is relatively small (29,799 users out of 500 million users as of June 26, 2012) but, given the nature of those users, it is clear that any sensible ranking algorithm should provide them the highest scores.

Therefore, using Twitter's API all the available verified accounts were obtained and checked against the collected dataset. Of course, not all of them were present in the dataset: some of the verified accounts are not tweeting in English and many of them did not exist at the moment of the data collection (2009). Nevertheless, 4,884 verified accounts were found in the dataset, a 0.3% of the total number of users; this is a relatively small number but it is highly significant given that only 30 thousand users from the 500 million user base in Twitter has got a verified account.

Using a likelihood-ratio test in a way analog to that of [29] we obtained a list of distinctive terms (see Table 1) from the biographies of verified accounts found in the dataset. As it can be seen, despite the presence of the "verified badge", it is pretty common to emphasize that the account is the official channel of communication in

[12] Twitter suspends accounts when the user is violating the terms of use; suspicious use associated with spamming is, in all probability, one of the most frequent reasons.

Twitter for that user. In addition to that, other phrases reveal that companies, news media outlets, politicians, artists and sport figures comprise the majority of verified accounts.

With regards to their tweeting behavior, Table 4 provides details and comparison with both spammers and average users; nevertheless, a summary of the most interesting features is provided here. As it was expected, verified accounts have huge follower bases and they follow thousands of accounts, much more than the average user and somewhat between spammers and aggressive marketers. They tweet much more than average users and, again, they are in between spammers and aggressive marketers in this regard. They use URLs much more than average users but much less than spammers. They use hashtags much less than average users, marketers and spammers which could suggest that verified accounts (i.e. celebrities) prefer not embrace with mass movements. Instead, they retweet much more than average users (although less than marketers) and get much involved in conversations than any other group of users; this could imply that they are really trying to personally interact with their followers.

| | | | |
|---|---|---|---|
| official | songwriter | nba | nbc |
| official twitter | new york times | comedian | mtv |
| news | cnn | dj | espn football |
| twitter page | on tour | tv | tips |
| band | host | olympic gold medalist | twitter channel |
| new album | tweets | producer | gov |
| twitter account | exclusive | follow us | washington |
| in stores now | fox news | nfl | spiegel online |
| espn | singer songwriter | official home | email |
| updates | correspondent | yahoo | the guardian |
| singer | recording artist | records | official tweets |
| congressional district | news commentary | governing body | fox sports |
| latest news | legal notice | television | grammy |
| twitter feed | anchor | world's leading | the leading |
| team | congressman | actress | rock band |

**Table 1.** The 60 most distinctive terms appearing in biographies of verified accounts.

## 4.3. Data preparation: abusive users

As we have described in the Research motivation section, in addition to obtain sensible rankings for relevant users, we are interested in the feasibility of rank prestige methods less vulnerable to linking malpractice[13]. This question is highly pertinent because one or more of such methods will certainly be used to rank users to find the most relevant/trustworthy content producers. In fact, at the moment of this writing Google seems to be already applying its *PageRank* method to rank Twitter users [47] in order to offer the most relevant tweets as search results.

We have already said that a desirable feature of ranking algorithms is that of demoting spammers. Hence, to compare different algorithms and test their respective performance we need a group of users who are actually trying to abuse Twitter linking. As many other so-called social services, Twitter is no immune to the spam problem. In addition to promoting dubious websites and products, Twitter spammers are also furiously engaged in getting followers; in fact, they have both more followees and followers than legitimate users. According to Yardi *et al.* (2010), they triple the number of both kinds of connections; these authors argue that "*spammers invest a lot of time following other users (and hoping other users follow them back)"*. An explanation we found highly plausible.

Whether this is done on purpose, anticipating the application of prestige algorithms to the user graph or, conversely, it is just a happy coincidence when spammers were just trying to find users to click the links they publish is irrelevant. The matter is that those spurious connections are to be treated by the eventual algorithm in the same way that legitimate links and, thus, spammers can obtain an undeserved authority within the user graph.

Thus, to collect the necessary data for the experiments, a method was needed to detect spammers. In Section 2 it was explained that machine learning methods are an option but they require a sample of both legitimate users and spammer for training and, on top of that, it was shown that the simple algorithm proposed by Yardi *et al.* provided a similar performance. Thus, such a method was chosen. The algorithm is based on the presence of URLs and a selection of keywords in the tweets, in addition to the matching of certain pattern in the user

[13] i.e. Creating links to earn reputation by exploiting knowledge about the algorithm instead of creating links as the naturally expected outcome of daily use of the service.

names[14]. Thus, we implemented an analogous version of their technique achieving similar performance: 87.32% precision versus the 91% reported by them.

That spam detection system detected 9,369 spammers in our collection of tweets. By examining a representative sample we found that about 24% of those users were already suspended by Twitter[15], 21% of them were promoting *"money making"* techniques, about 11% were *"copywriters"* (i.e. they replicate content from other users, RSS feeds or publish plagiarized websites), 8% promote methods to get followers and/or website traffic, and the rest of them are a mixture of self-proclaimed experts in SEO, marketing, weight loss, etc.

In a similar way to that of verified accounts, spammers biographies were mined to find distinctive terms (see Table 2 for a more exhaustive list). As it was expected, terms such as *business*, *money*, *internet marketing*, *social media* and *SEO* were at the top of the list. It must be noticed that those terms are not only popular among spammers but among other Twitter users. As [55] said of them *"[they] tread a fine line between reasonable and spam-like behavior in their efforts to gather followers and attention"*. We will denote those users as *aggressive marketers* and, thus, we decided to expand the group of abusive users from pure spammers to also include those marketers[16].

To find them we prepared a list of terms commonly occurring in spammer biographies which were also frequently associated with marketers and other heavy-following users. By doing so, we found another 22,290 users which cannot be labeled as spammers but, we thought, could exhibit some abusive behavior with regards to connecting to other users. Table 3 shows a list of the 60 most distinctive terms for these aggressive marketers; please notice the degree of coincidence with spammers.

Of course, a mere similarity in biographies is not evidence of aggressive marketers trying to abuse Twitter. Because of that, we analyzed several characteristics of tweeting behavior for both spammers and aggressive marketers, and compared them to those of average Twitter users and verified accounts. Table 4 contains all the details and, thus, we will just summarize the most interesting findings.

First, spammers have both much more followers and followees than average users (although less than verified accounts). This is consistent with the findings of Yardi *et al.* However, in our datasets spammers do not triple those numbers with regards to common users, but they multiply them by almost 40! Aggressive marketers are in between but much closer to spammers than to the average twitterer: they have about 15 times the number of connections than an average user.

With regards to the number of tweets published, aggressive marketers double the frequency of average users and spammers publish 7 times the number of tweets an average user does. This probably means that spammers, in contrast to marketers, are publishing tweets in an automatic fashion.

Regarding different features of the tweets themselves there are some important differences between spammers, marketers, and average users. Firstly, virtually every tweet published by a spammer contains a URL (90%); marketers use URLs in one in three tweets, while average users tend to use URLs in about one in five tweets. Secondly, both marketers and spammers employ hashtags more than average users but the differences, although substantial, are not as pronounced as with other features. Surprisingly, the number of hashtags include by these different groups is mostly the same on average. Lastly, one feature that again highlights the robot nature of most spammers is the much lower level of retweeting, in particular, and conversations they get involved, in general. As it was expected, marketers are much more prone to retweet than average users (two times) and also get much more involved in conversations than them.

To sum up, we have described two similar, albeit not identical, groups of abusive users –namely spammers and marketers– which exhibit several features very different from those of average users and verified accounts. Both have further more followers and followees than the average user, both publish more than average users, and both promote URLs more than average users. Besides, as we have stated above, the line between aggressive marketers and spammers is not always totally clear. In fact, one could assume that relatively few users will respond to the stereotypical spammer profile and, to the contrary, many users would exhibit a more or less marked spamming behavior.

---

[14] Kwak et al. (2010) employed a different mechanism: labeling as spam those tweets including one or more *trending topics*. In addition to require additional knowledge (i.e. knowing which are the trending topics at a given time) there is also another problem with this approach: as we will show later, the average number of hashtags (usually appearing in trending topics) is well down below 3, even for spammers.

[15] Needless to say, one of the main reasons for account suspension in Twitter is spamming behavior.

[16] It must be noted that Lee, Caverlee and Webb [34] considered aggressive marketers a subclass of spammers; they called them "promoters".

Anyway, and spite of being an oversimplification, for the purpose of evaluating the different graph centrality algorithms with regards to spammers demotion we are going to consider a single group spammers-marketers assuming that their linking behavior, as a whole, is trying to abuse, or at least cheat, the assumptions underlying relationships in Twitter: i.e. that users should follow one another when they are genuinely interested in what that the other is publishing. Thus, in the following sections we are going to analyze in which way the global authority/reputation is divided among the different groups (verified accounts, spammers-marketers and the rest of users), and which algorithms are less prone to be abused while, at the same time, providing a better ranking of relevant users within the social network.

| | | | |
|---|---|---|---|
| marketing | free | expert | online marketer |
| internet | help | investor | weight loss |
| marketer | deals | people | trump network |
| online | make money | network marketing | helping others |
| business | real estate | mlm | media marketing |
| money | forex | blog | marketing coach |
| social | coach | traffic | money making |
| internet marketer | home | success | help people |
| internet marketing | real | online marketing | forex trading |
| social media | news | network marketer | helping people |
| entrepreneur | money online | affiliate marketer | home based |
| affiliate | helping | making money | home business |
| network | tips | online business | internet entrepreneur |
| media | affiliate marketing | estate investor | forex trader |
| seo | web | small business | business coach |

**Table 2.** The 60 most distinctive terms appearing in Twitter spammers biographies.

| | | | |
|---|---|---|---|
| **entrepreneur** | consultant | small | owner |
| **marketing** | **social** | others | **home** |
| **internet** | **people** | **helping others** | **affiliate marketing** |
| **real estate** | **helping people** | **success** | **online marketer** |
| estate | **media** | **media marketing** | follow |
| **real** | **social media** | **help people** | business owner |
| **marketer** | **affiliate** | entreprenuer *(sic)* | speaker |
| **online** | marketing consultant | broker | guru |
| **business** | **web** | realtor | estate broker |
| **internet marketing** | **small business** | **estate investor** | search |
| **helping** | **coach** | **network marketing** | sem |
| **internet marketer** | **help** | estate agent | ppc |
| **seo** | **network** | agent | successful |
| **money** | **free** | **make money** | blogger |
| **online marketing** | **investor** | **expert** | **network marketer** |

**Table 3.** The 60 most distinctive terms appearing in aggressive marketers biographies. Those terms in bold also appear in the top-60 list for spammers.

| | Verified accounts | Spammers | Aggressive marketers | Average user |
|---|---|---|---|---|
| **Avg. in-degree** | 34784.36 | 3203.28 | 1338.83 | 82.36 |
| **Avg. out-degree** | 1412.80 | 3156.09 | 1245.35 | 82.36 |
| **Avg. # of tweets over the whole period and SD** | 15.29 (32.74) | 41.25 (80.99) | 12.93 (34.07) | 5.60 (19.45) |
| **% of tweets including URLs** | 31.11% | 90.42% | 32.86% | 18.21% |
| **Avg. # of URLs per tweet including URLs** | 1.013 | 1.018 | 1.015 | 1.014 |
| **% of tweets including hashtags** | 7.01% | 11.54% | 8.83% | 7.98% |
| **Avg. # of tags per tweet including hashtags** | 1.25 | 1.41 | 1.42 | 1.50 |
| **% of retweets over total tweets** | 4.87% | 2.97% | 6.50% | 2.87% |
| **% of "conversations" over total (excluding retweets)** | 32.88% | 6.86% | 21.48% | 19.26% |
| **Avg. # of users referred in conversational tweets (excluding retweets)** | 1.13 | 1.17 | 1.13 | 1.09 |

**Table 4.** Features characterizing verified accounts, spammers, aggressive marketers, and average users.

### 4.4. Evaluation method

As we have already said we are not assuming any prior "correct" ranking for the users; we consider user rankings just a tool to find the most relevant source of information at a given time and, thus, a ranking algorithm will be judged by its ability to rank atop relevant users as a whole while avoiding abusive users achieving undeserved rankings.

Hence, the evaluation process is quite straightforward. All of the different methods were applied to the Twitter graph to obtain a user ranking. Then, we compared the positions reached by spammers and marketers on one side, and verified accounts on another side. The lower the rankings abusive users reach while the higher for verified accounts, the better the method is.

In the following section we provide the minimum, average, and median positions for the different user classes across different deciles. Such numbers will help to understand the positioning of both verified accounts and abusive users in relation to the rest of users in the social network. In addition to that, we will graphically show the percentage of verified accounts atop the ranks, the percentage of abusive users found as one moves down the ranking, and the level of agreement between the different rankings.

## 5. Results

### 5.1. Prestige of abusive users and verified accounts when applying *PageRank*

As it has been aforementioned, 4,884 verified accounts appear in the collected dataset. They amount for a mere 0.3% of the users in the graph but, still, they accumulate 12.7% of the total *PageRank* in the graph. Regarding their positioning in the global ranking (see Table 5, and Figures 3 and 4), 85% of the verified accounts appear among the top 10% of users.

With regards to the subset of abusive users, about 50% of the spammers detected in the collection of tweets did not appear in the graph[17]. Those who are present account for 0.25% of the users but they agglutinate 1.4% of the total *PageRank* in the graph. Regarding the aggressive marketers, 98% of them appear in the graph accounting for 3.3% of the total *PageRank*. The acute difference from spammer to marketer presence in the graph gives an idea of the work devoted by Twitter to get rid of spammers.

Thus, the whole set of spammers and marketers which represent a mere 1.5% of the users manage to grab 4.7% of the available *PageRank*.

With regards to their positioning in the global ranking (see Figures 5 and 6), 90% of spammers are, approximately, among the 60% of top ranked users and, in fact, half of the spammers appear well above the top 10% of Twitter users. On the other hand, 90% of the aggressive marketers are among the 80% of top ranked users and half of them appear above the top 20% of users.

### 5.2. Prestige of abusive users and verified accounts when applying *HITS*

When applying *HITS* to the Twitter user graph verified accounts grab 1.65% of the available authority rank while spammers grab 5.20% and aggressive marketers account for another 11%. Thus, a reduce subset of users (1.5% of them) which are on top of that abusive own 16.20% of the global authority outscoring verified accounts (i.e. relevant users).

Both spammers and marketers are pretty good positioned (see Table 6). Half of the spammers appear at the top 10% of positions and 90% of them are among the 40% better situated users. Aggressive marketers' situation is not as good but equally impressive: half of them appear at the top 20% positions, and 90% of them are among the 60% better positioned users.

With regards to verified accounts, almost half of them appear at the top 10% positions and 90% of them are among the top 40% better ranked users. This, however, is not very different of the rankings achieved by spammers and marketers.

[17] Let's remember that the user graph was crawled once the collection of tweets was completed.

## 5.3. Prestige of abusive users and verified accounts when applying *NodeRanking*

When ranking users according to *NodeRanking*, verified accounts reach 11.94% of the global available prestige while spammers and aggressive marketers account for 1.62% and 3.86%, respectively. The amounts reached by abusive users are about 15% larger than those achieved when using *PageRank* while the prestige grabbed by verified accounts is slightly lower.

With regards to their positioning (see Table 7), 90% of verified accounts appear near the top 15% users, and half of them are among the top 2% of users. With regards to abusive users, 90% of the spammers are among the 60% best situated users and half of them appear at the top 10% positions. 90% of the aggressive marketers are above 30% of the users and half of them are among the top 20% users. Such results are pretty similar to those obtained by applying standard *PageRank*.

## 5.4. Prestige of abusive users and verified accounts when applying *TunkRank*

When ranking Twitter users according to *TunkRank*[18], verified accounts grab 21.47% of the available global prestige while spammers and aggressive marketers account for 0.74% and 1.94%, respectively. Hence, the amount grabbed by abusive users is roughly half the one obtained when applying *PageRank* while the amount reached by verified accounts is 69% larger.

Attending to their positioning (see Table 8), 90% of verified accounts are at the top 15%, and half of them are at the top 1%. In contrast, 90% of the spammers are among the 70% of best positioned users, and half of them appear above the top 20%. Regarding aggressive marketers, there are no great differences between them and common users, although half of them are above the 40% top positioned users.

## 5.5. Prestige of abusive users and verified accounts when applying *TwitterRank*

When ranking Twitter users according to *TwitterRank*, verified accounts reach 9.12% of the available global prestige. In contrast, spammers and aggressive marketers account for 0.0003% and 0.00025%, respectively. In other words, using *TwitterRank*, both groups of abusive users reach a virtually negligible prestige (although, again, spammers manage to outperform marketers).

With regards to their positioning (see Table 9), 90% of verified accounts are roughly among the top 60% users, and half of them are among the top 20%. Surprisingly, 90% of the spammers are among the top 30% users and half of them appear among the 10% best positioned users. Aggressive marketers, on the other hand, seem to be slightly better positioned than average users.

The reason for these apparently contradictory results (namely, the impressive prestige reduction for spammers while still achieving top positions) is that *TwitterRank* distributes the prestige in a highly biased way: in fact, top 10 users account for 77% of the prestige and top 25 users for 95.5%[19]. That is, virtually all of the users in the network achieve no prestige at all and, in spite of that, spammers manage to be "one-eyed kings in the land of the blind" even outperforming a large number of verified accounts.

## 5.6. Prestige of abusive users and verified accounts when applying *PageRank* with paradoxical discounting

When applying *PageRank* with paradoxical discounting to the Twitter user graph, verified accounts grab 10.87% of the global *PageRank* while spammers can only grab 0.22% and aggressive marketers account for 1.05%. Thus, with regards to standard *PageRank*, spammers loose -84.3%, and marketers -68.2%.

One of the consequences of applying paradoxical discounting to *PageRank* is that many users reach a *PageRank* which is virtually zero and, hence, all of those users tie for the last position (see Table 10). 40.2% of the spammers end up in that bin while the rest of them appear among the 30% top ranked users. With regards to aggressive marketers, 55% of them reach a negligible *PageRank* value but the remaining 45% is among the top 30% users in the graph. Verified accounts behave as expected with 90% of them among the top 14% of users and half of them above the 2%. These somewhat mixed results are later discussed.

---

[18] As it was explained before, *TunkRank* requires a constant $p$ which is the probability of users retweeting. For this experiment we employed a value of 2.87% which was found during the analysis of the dataset (see Table 4). As a side note, the value employed by `tunkrank.com` is 5%.

[19] Let's compare, for instance, with *PageRank* which distributes only 1.2% and 2.3% of the available prestige to top 10 and top 25 users, respectively.

### 5.7. Prestige of abusive users and verified accounts when applying *PageRank* to a "pruned" user graph

As it was described in a previous section, paradoxical discounting could be used to "prune" the graph which would, in turn, be ranked by means of standard *PageRank*. When applying this approach to the Twitter graph, verified accounts grab 10.98% of the available *PageRank*, spammers 1.84%, and aggressive marketers account for 4.27%. Regarding of their positioning (see Table 11), 90% of verified accounts are close to the top 15% of users, and half of them are above the top 2%. With regards to abusive users, 90% of the spammers are best positioned than half of the users, and half of them are among the top 10% users; aggressive marketers, 90% of them appear among the 70% best positioned users, and half of them are among the top 20% users. These results will be later discussed because of their implications regarding reciprocal linking in Twitter.

| Decile | All users | | | Verified accounts | | | Spammers | | | Aggressive marketers | | |
|---|---|---|---|---|---|---|---|---|---|---|---|---|
| | **Min.** | **Avg.** | **Median** | **Min.** | **Avg.** | **Median** | **Min.** | **Avg.** | **Median** | **Min.** | **Avg.** | **Median** |
| 9th | 165,301 | 85,645.5 | 82,651 | 1 | 187.5 | 178 | 17,940.5 | 9,686.5 | 9,619 | 40,265 | 20,668.7 | 20,661.5 |
| 8th | 330,606.5 | 165,308.5 | 165,301 | 422 | 1,115.4 | 1,069 | 32,562 | 17,148 | 17,940.5 | 86,683 | 41,880.3 | 40,267.5 |
| 7th | 495,985.5 | 247,962.5 | 330,606.5 | 1,989 | 3,325 | 3,292 | 54,935 | 25,679.8 | 24,227.5 | 146,608 | 66,403.4 | 62,855 |
| 6th | 661,268 | 330,616.5 | 330,606.5 | 4,846 | 6,936.5 | 6,945 | 81,251 | 36,148.6 | 32,562 | 221,907 | 95,327.3 | 86,683 |
| **5th** | **826,734.5** | **413,270.5** | **413,256.5** | 9,351 | 12,723.5 | 12,438 | **117,859** | **48,788.8** | **42,085.5** | **318,219** | **129,933.9** | **115,519** |
| 4th | 992,189 | 495,924 | 495,985.5 | 16,956 | 23,562.6 | 23,000 | 176,590 | 69,949.4 | 54,935 | 446,082 | 171,494.7 | 146,608 |
| 3rd | 1,156,984 | 578,578.1 | 578,550.5 | 31,638 | 41,834.2 | 41,204 | 282,436.5 | 87,641.5 | 66,906 | 612,565 | 222,296.9 | 180,888.5 |
| 2nd | 1,323,166 | 661,232 | 661,268 | 54,562 | 73,861.5 | 73,066 | 458,004 | 121,866.6 | 81,251 | 846,257 | 284,973.9 | 221,907 |
| **1st** | **1,487,118.5** | **743,886.1** | **743,738.5** | 97,461 | 130,283.4 | 130,608 | **846,790.5** | **178,970.1** | **98,665.5** | **1,175,085.5** | **364,625.6** | **267,299.5** |

**Table 5.** Minimum, average, and median position for Twitter users, verified accounts, spammers, and marketers, across different deciles when using standard *PageRank* to rank the Twitter user graph.

| Decile | All users | | | Verified accounts | | | Spammers | | | Aggressive marketers | | |
|---|---|---|---|---|---|---|---|---|---|---|---|---|
| | **Min.** | **Avg.** | **Median** | **Min.** | **Avg.** | **Median** | **Min.** | **Avg.** | **Median** | **Min.** | **Avg.** | **Median** |
| 9th | 165,303 | 82,654.5 | 82,656.5 | 13 | 5,737 | 5,401 | 4,032 | 2,155.7 | 2,215 | 14,107 | 6,602.6 | 6,330 |
| 8th | 330,598 | 165,308.5 | 165,303 | 14,584 | 35,797.8 | 36,430 | 9,121 | 4,243.3 | 4,032 | 35,548 | 15,271 | 14,126 |
| 7th | 496,058 | 247,962.5 | 247,978.5 | 54,186 | 71,102.1 | 71,630 | 16,608.5 | 6,986.2 | 6,131 | 66,727 | 26,847.7 | 23,713 |
| 6th | 661,302.5 | 330,616.6 | 330,598 | 87,671 | 104,383.3 | 104,069 | 27,873 | 10,599.9 | 9,121 | 109,830.5 | 42,021 | 35,548 |
| **5th** | **826,483** | **413,270.7** | **413,221** | **121,819** | **140,075.5** | **138,731** | **45,216** | **15,626.5** | **12,360** | **170,056.5** | **61,174.6** | **49,404** |
| 4th | 990,992 | 495,924.2 | 496,058 | 159,859 | 181,858.8 | 181,941 | 77,618.5 | 22,933.7 | 16,608.5 | 248,962 | 85,539.6 | 66,727 |
| 3rd | 1,156,625 | 578,579.5 | 578,467.5 | 206,551 | 235,986.8 | 236,208 | 132,640.5 | 34,292.9 | 21,127 | 373,174 | 116,890.2 | 87,372 |
| 2nd | 1,320,297 | 661,235.9 | 661,302.5 | 269,506 | 310,013 | 308,410 | 252,771.5 | 53,755.8 | 27,873 | 574,641 | 160,334.7 | 109,830.5 |
| **1st** | **1,490,091** | **743,952.4** | **742,973.5** | **361,512** | **433,174.1** | **434,604** | **518,039.5** | **88,726.6** | **35,092.5** | **915,979.5** | **223,959.3** | **136,489** |

**Table 6.** Minimum, average, and median position for Twitter users, verified accounts, spammers, and marketers, across different deciles when using *HITS* to rank the Twitter user graph.

| Decile | All users | | | Verified accounts | | | Spammers | | | Aggressive marketers | | |
|---|---|---|---|---|---|---|---|---|---|---|---|---|
| | **Min.** | **Avg.** | **Median** | **Min.** | **Avg.** | **Median** | **Min.** | **Avg.** | **Median** | **Min.** | **Avg.** | **Median** |
| 9th | 165,307.5 | 82,654.5 | 82,654.5 | 1 | 187 | 174 | 16,214 | 8,724.1 | 8,917 | 37,274.5 | 19,002 | 18,822 |
| 8th | 330,602.5 | 165,308.5 | 165,307.5 | 433 | 1,213.8 | 1,151 | 30,166 | 15,615.7 | 16,214 | 81,564 | 39,015.7 | 37,294 |
| 7th | 495,905 | 247,962.5 | 247,953.5 | 2,238 | 3,654.9 | 3,603 | 51,140.5 | 23,685.4 | 22,262 | 139,706 | 62,453.7 | 58,977 |
| 6th | 661,365 | 330,616.5 | 330,602.5 | 5,322 | 7,712.3 | 7,611 | 75,816.5 | 33,459.1 | 30,166 | 212,520.5 | 90,346.8 | 81,564 |
| **5th** | **826,287** | **413,270.5** | **413,309.5** | **10,493** | **14,611.2** | **14,502** | **111,405** | **45,443.1** | **39,187** | **306,308.5** | **123,786.1** | **108,975** |
| 4th | 992,190 | 495,924 | 495,905 | 19,453 | 27,196.2 | 26,826 | 168,385 | 60,910.7 | 51,140.5 | 429,445.5 | 163,948.5 | 139,706 |
| 3rd | 1,175,410.5 | 578,578.1 | 578,615.5 | 35,956 | 48,267.2 | 48,183 | 276,355.5 | 83,027.3 | 62,142 | 585,996 | 212,707 | 172,703.5 |
| 2nd | 1,322,298 | 661,232.2 | 661,365 | 62,444 | 83,683.9 | 82,489 | 448,658.5 | 116,781.9 | 75,816.5 | 808,880 | 272,550.8 | 212,520.5 |
| **1st** | **1,487,014.5** | **743,886.2** | **743,947** | **109,264** | **144,141.3** | **142,370** | **834,146** | **172,969.8** | **93,348.5** | **1,136,582** | **349,228.7** | **256,175.5** |

**Table 7.** Minimum, average, and median position for Twitter users, verified accounts, spammers, and marketers, across different deciles when using *NodeRanking* to rank the Twitter user graph.

| Decile | All users | | | Verified accounts | | | Spammers | | | Aggressive marketers | | |
|---|---|---|---|---|---|---|---|---|---|---|---|---|
| | Min. | Avg. | Median | Min. | Avg. | Median | Min. | Avg. | Median | Min. | Avg. | Median |
| 9th | 165,313.5 | 82,654.5 | 82,651.5 | 1 | 168.4 | 165 | 29,875 | 15,806 | 15,428 | 70,623.5 | 36,099.8 | 36,341.5 |
| 8th | 330,583.5 | 165,308.5 | 165,313.5 | 356 | 889.8 | 879 | 68,423.5 | 32,130.3 | 29,875 | 155,965.5 | 73,447 | 70,667.5 |
| 7th | 495,982.5 | 247,962.5 | 247,988 | 1,513 | 2,367.6 | 2,365 | 114,073 | 51,371.2 | 48,712.5 | 270,273 | 118,997.9 | 110,180.5 |
| 6th | 661,167.5 | 330,616.5 | 330,583.5 | 3,329 | 4,512.6 | 4,474 | 187,660 | 75,544 | 68,423.5 | 401,255 | 173,802 | 155,965.5 |
| **5th** | **826,476.5** | **413,270.5** | **413,311** | **5,846** | **7,687.2** | **7,657** | **297,690.5** | **108,306.6** | **89,344.5** | **544,777** | **233,042.9** | **208,035.5** |
| 4th | 991,747.5 | 495,924 | 495,982.5 | 9,898 | 12,701.5 | 12,464 | 402,848.5 | 147,929.6 | 114,073 | 716,496 | 298,732.6 | 270,273 |
| 3rd | 1,157,027.5 | 578,578 | 578,592.5 | 16,479 | 21,506.1 | 21,331 | 572,314 | 195,811.7 | 146,916.5 | 901,781.5 | 371,047.8 | 337,262 |
| 2nd | 1,322,839 | 661,232.2 | 661,167.5 | 27,711 | 37,395.6 | 36,531 | 787,330.5 | 256,066.8 | 187,660 | 1,105,147 | 449,876.8 | 401,255 |
| **1st** | **1,487,335** | **743,886.1** | **743,963** | **51,026** | **77,552.1** | **74,914** | **1,096,915** | **330,910.7** | **237,916.5** | **1,332,358** | **535,239.9** | **467,271.5** |

**Table 8.** Minimum, average, and median position for Twitter users, verified accounts, spammers, and marketers, across different deciles when using *TunkRank* to rank the Twitter user graph.

| Decile | All users | | | Verified accounts | | | Spammers | | | Aggressive marketers | | |
|---|---|---|---|---|---|---|---|---|---|---|---|---|
| | Min. | Avg. | Median | Min. | Avg. | Median | Min. | Avg. | Median | Min. | Avg. | Median |
| 9th | 130,344.5 | 83,297.9 | 76,499.5 | 58 | 1,148.7 | 1,028 | 24,019.5 | 12,713.8 | 12,471 | 130,344.5 | 53,960.5 | 52,065.5 |
| 8th | 334,540.5 | 207,115.7 | 130,344.5 | 2,796 | 11,377 | 10,648 | 52,065.5 | 24,371.1 | 24,019.5 | 334,540.5 | 146,980.8 | 130,344.5 |
| 7th | 334,540.5 | 249,590.7 | 334,540.5 | 22,411 | 40,942.5 | 39,699 | 76,499.5 | 37,128.1 | 38,463.5 | 334,540.5 | 209,491.2 | 334,540.5 |
| 6th | 1,076,966 | 450,901.4 | 334,540.5 | 64,618 | 95,312.8 | 95,597 | 130,344.5 | 50,195.6 | 52,065.5 | 1,076,966 | 287,363.9 | 334,540.5 |
| **5th** | **1,076,966** | **576,114.3** | **334,540.5** | **131,440** | **172,739.2** | **170,968** | **130,344.5** | **66,239.5** | **52,065.5** | **1,076,966** | **445,255.4** | **334,540.5** |
| 4th | 1,076,966 | 659,589.2 | 334,540.5 | 220,912 | 276,859.1 | 274,768 | 130,344.5 | 76,911.9 | 76,499.5 | 1,076,966 | 550,524.5 | 334,540.5 |
| 3rd | 1,076,966 | 719,214.5 | 1,076,966 | 341,882 | 407,691.9 | 409,117 | 334,540.5 | 113,043.3 | 76,499.5 | 1,076,966 | 625,750 | 334,540.5 |
| 2nd | 1,076,966 | 763,933.5 | 1,076,966 | 492,637 | 575,699.9 | 566,181 | 334,540.5 | 140,699.9 | 130,344.5 | 1,076,966 | 682,142.4 | 1,076,966 |
| **1st** | **1,076,966** | **798,714.9** | **1,076,966** | **653,951** | **785,413.7** | **750,574** | **334,540.5** | **162,258.9** | **130,344.5** | **1,076,966** | **726,022.8** | **1,076,966** |

**Table 9.** Minimum, average, and median position for Twitter users, verified accounts, spammers, and marketers, across different deciles when using *TwitterRank* to rank the Twitter user graph.

| Decile | All users | | | Verified accounts | | | Spammers | | | Aggressive marketers | | |
|---|---|---|---|---|---|---|---|---|---|---|---|---|
| | Min. | Avg. | Median | Min. | Avg. | Median | Min. | Avg. | Median | Min. | Avg. | Median |
| 9th | 134,298 | 83,292.2 | 85,303 | 5 | 343.4 | 346 | 49,793 | 26,242.4 | 25,470 | 85,303 | 48,189 | 49,793 |
| 8th | 312,904.5 | 196,352.9 | 134,298 | 643 | 1,482.5 | 1,431 | 134,298 | 52,009.5 | 49,793 | 312,904.5 | 92,727.9 | 85,303 |
| 7th | 1,055,174.5 | 293,060 | 312,904.5 | 2,457 | 3,699.8 | 3,656 | 134,298 | 79,459,2 | 85,303 | 312,904.5 | 166,108.9 | 134,298 |
| 6th | 1,055,174.5 | 483,588.6 | 312,904.5 | 5,077 | 6,770 | 6,606 | 312,904.5 | 136,279.2 | 134,298 | 312,904.5 | 202,816.2 | 312,904.5 |
| **5th** | **1,055,174.5** | **597,905.8** | **312,904.5** | **8,806** | **11,429.1** | **11,403** | **312,904.5** | **171,635.4** | **134,298** | **1,055,174.5** | **298,771.7** | **312,904.5** |
| 4th | 1,055,174.5 | 674,116.8 | 1,055,174.5 | 14,320 | 18,252.7 | 18,239 | 1,055,174.5 | 197,063.9 | 134,298 | 1,055,174.5 | 424,819.6 | 312,904.5 |
| 3rd | 1,055,174.5 | 728,553.7 | 1,055,174.5 | 23,048 | 29,889.2 | 29,454 | 1,055,174.5 | 319,766.9 | 312,904.5 | 1,055,174.5 | 514,893.8 | 312,904.5 |
| 2nd | 1,055,174.5 | 171,635.4 | 1,055,174.5 | 38,618 | 51,745.5 | 50,453 | 1,055,174.5 | 411,591.5 | 312,904.5 | 1,055,174.5 | 582,417.3 | 312,904.5 |
| **1st** | **1,055,174.5** | **197,063.9** | **1,055,174.5** | **69,525** | **100,689.4** | **97,890** | **1,055,174.5** | **483,170.8** | **312,904.5** | **1,055,174.5** | **634,959.2** | **312,904.5** |

**Table 10.** Minimum, average, and median position for Twitter users, verified accounts, spammers, and marketers, across different deciles when using *PageRank* with paradoxical de-weighting to rank the Twitter user graph.

| Decile | All users | | | Verified accounts | | | Spammers | | | Aggressive marketers | | |
|---|---|---|---|---|---|---|---|---|---|---|---|---|
| | **Min.** | **Avg.** | **Median** | **Min.** | **Avg.** | **Median** | **Min.** | **Avg.** | **Median** | **Min.** | **Avg.** | **Median** |
| 9th | 165,313.5 | 82,654.5 | 82,653.5 | 1 | 203.4 | 186 | 14,720 | 7,856.4 | 8,083 | 34,783 | 17,640.9 | 17,405.5 |
| 8th | 330,623 | 165,308.5 | 165,313.5 | 480 | 1,341.4 | 1,244 | 27,032.5 | 14,024.2 | 14,720 | 76,738.5 | 36,527.7 | 34,790.5 |
| 7th | 495,918.5 | 247,962.5 | 247,965 | 2,484 | 4,029.7 | 3,896 | 46,741 | 21,430.7 | 19,863.5 | 133,149 | 58,984.4 | 55,293.5 |
| 6th | 661,199 | 330,616.5 | 330,623 | 5,903 | 8,493.6 | 8,304 | 69,424.5 | 30,400.2 | 27,032.5 | 203,123.5 | 85,750.6 | 76,738.5 |
| **5th** | **826,346** | **413,270.5** | **413,236.5** | **11,624** | **16,203.7** | **16,167** | **102,791.5** | **41,513.2** | **35,646** | **294,846** | **117,985.8** | **103,986.5** |
| 4th | 991,809.5 | 495,924.1 | 495,918.5 | 21,604 | 29,449.8 | 28,972 | 155,756 | 55,942.5 | 46,741 | 412,074.5 | 156,724.4 | 133,149 |
| 3rd | 1,156,691 | 578,578.1 | 578,616.5 | 38,474 | 51,076.5 | 50,115 | 251,543 | 76,324.3 | 56,803.5 | 564,040 | 203,651.2 | 165,030 |
| 2nd | 1,322,537 | 661,232.3 | 661,199 | 66,758 | 87,946.3 | 88,293 | 416,371 | 107,509.6 | 69,424.5 | 775,863 | 261,087.1 | 203,123.5 |
| **1st** | **1,486,604.5** | **743,886** | **743,962.5** | **112,303** | **146,162.1** | **144,613** | **757,974** | **159,341.3** | **85,980** | **1,097,854.5** | **334,896.5** | **244,943.5** |

**Table 11.** Minimum, average, and median position for Twitter users, verified accounts, spammers, and marketers, across different deciles when using *PageRank* to rank a Twitter user graph "pruned" by means of paradoxical de-weighting.

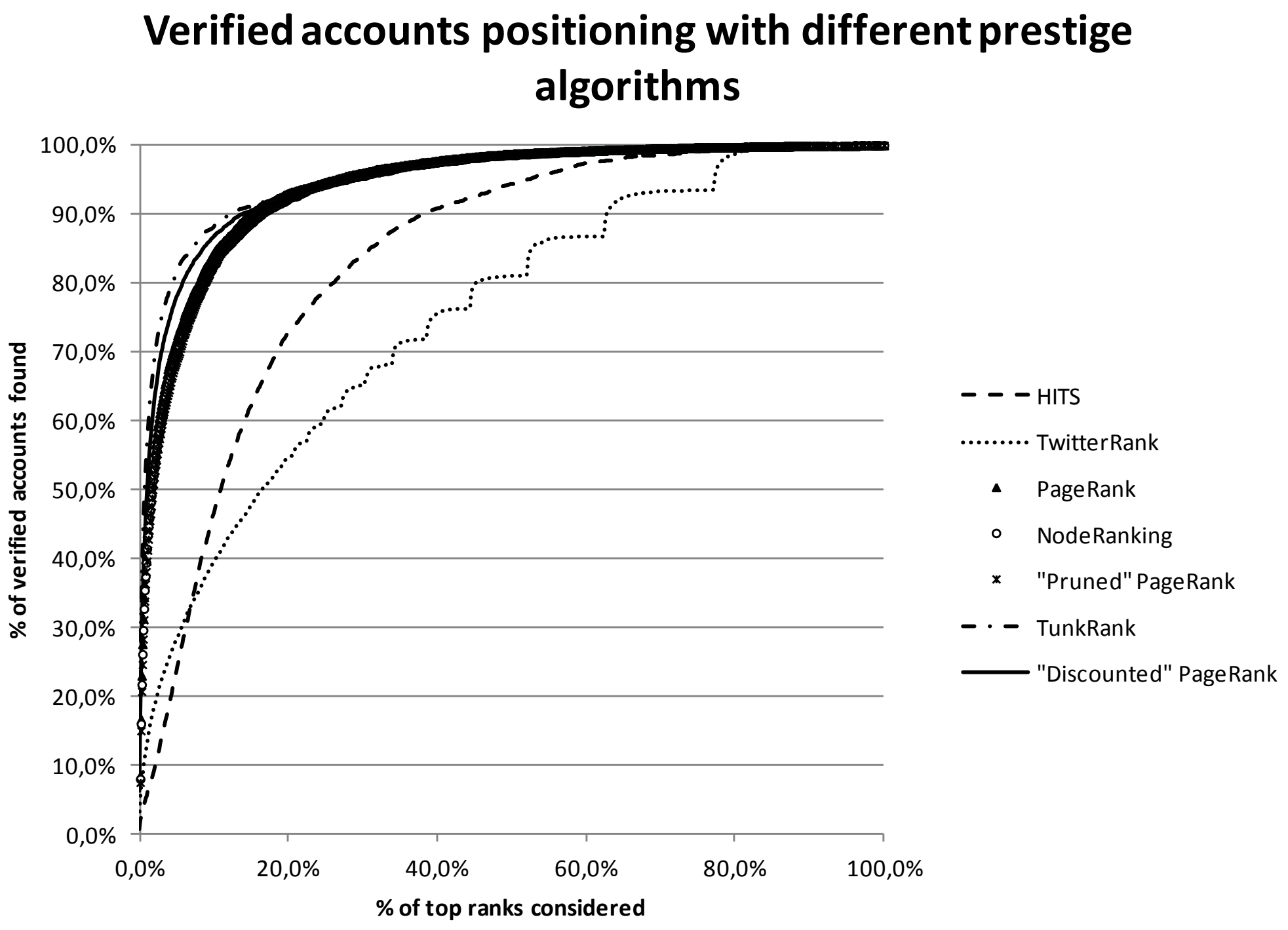

**Figure 3.** Percent of verified accounts found for different slices of the users ranking.

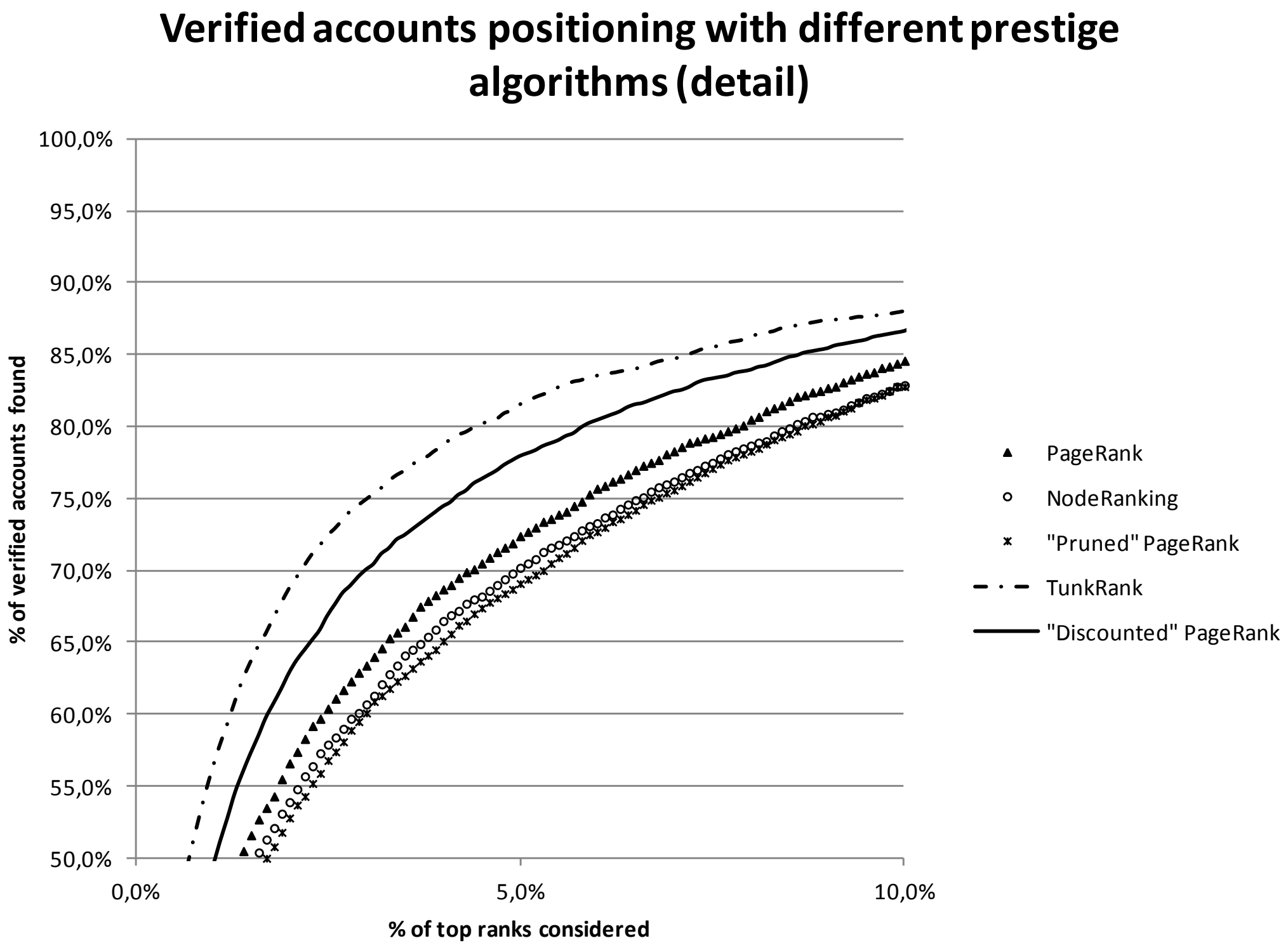

**Figure 4.** Details showing the differences among the best performing algorithms when ranking verified accounts.

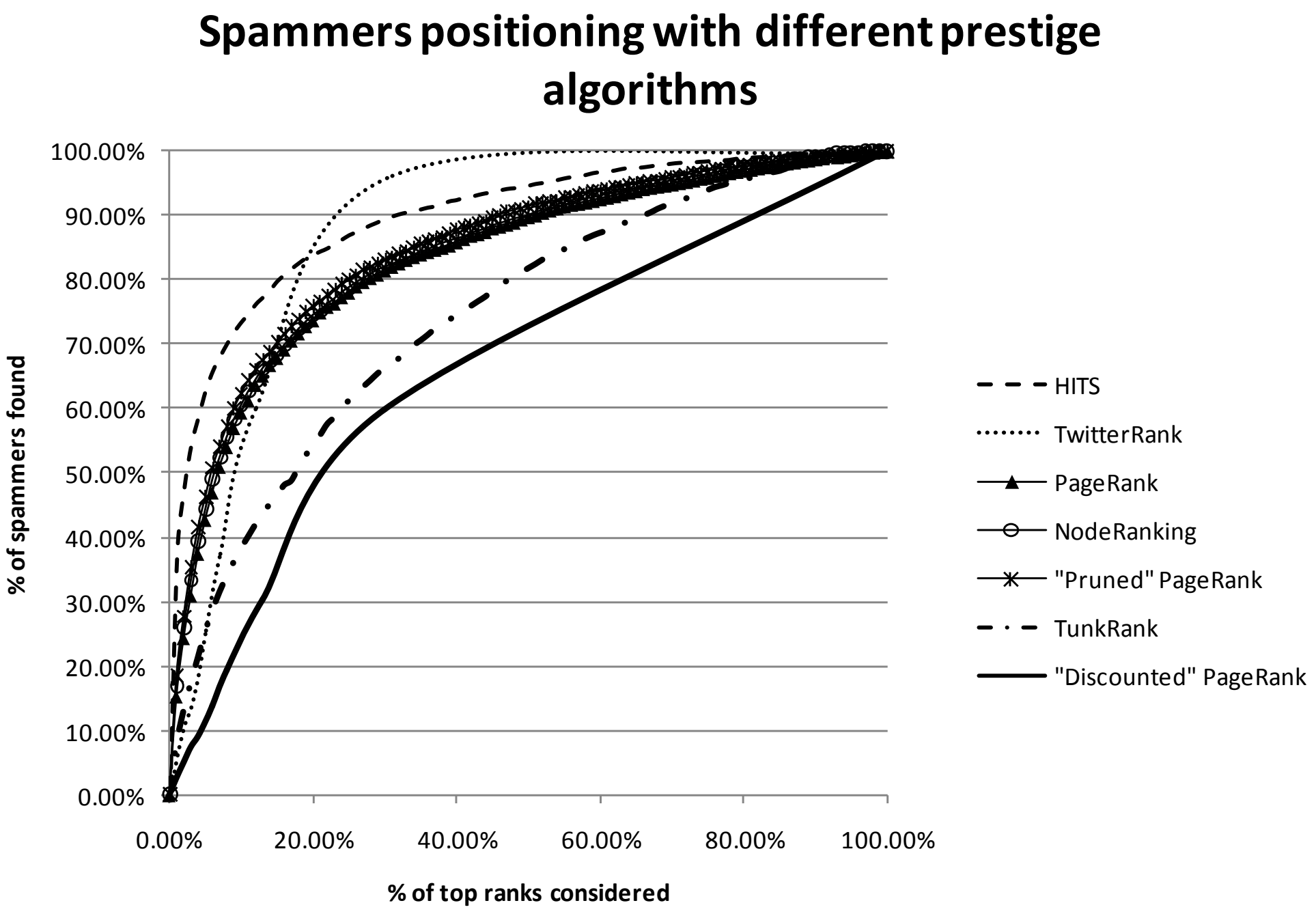

**Figure 5.** Percent of spammers found for different slices of the users ranking.

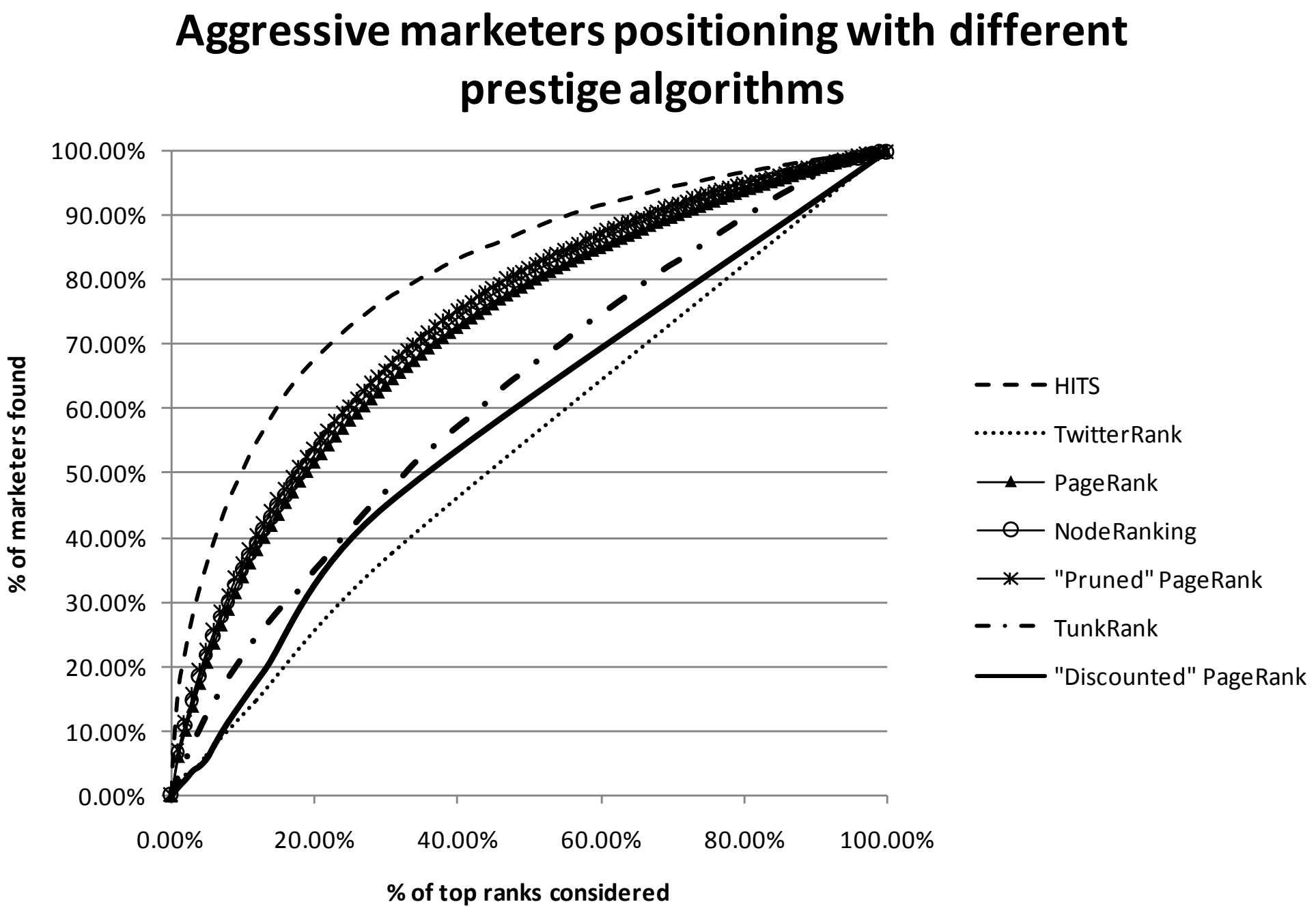

**Figure 6.** Percent of aggressive marketers found for different slices of the users ranking.

## 5.8. Agreement between the different rankings

Up to now we have shown the ability of the different algorithms to "penalize" abusive users while still ranking atop relevant users (i.e. verified accounts). In addition to that, it would be interesting to check if the induced rankings are "plausible" and the level of agreement between them. Table 12 shows the top 30 users according to the different ranking algorithms.

As it can be seen, *PageRank*, *NodeRanking*, *TunkRank* and "pruned" *PageRank* exhibit a large level of agreement; all of them highly rank celebrities and personalities, news wires, and a few companies. "Discounted"

*PageRank*, promotes several new users to the top rank, most of them musicians or related to alternative news wires. All of those algorithms spotted plenty of verified accounts among the top 30 users.

*HITS* and *TwitterRank* are another question. *HITS* top rank is plagued with self-proclaimed entrepreneurs, CEOs, marketers and gurus. *TwitterRank*, probably because of the importance of content similarity between different users, promotes to the top of the list mainly feeds and robots interwoven in tight networks. The presence of verified accounts in both top 30 user lists is virtually null.

Table 12, however, just provides anecdotal evidence. In order to gain details on the behavior of the different ranking algorithms we compared them by means of the normalized version of Kendall distance with a zero penalty parameter ([38] and [14], respectively). Figures 7 to 9 show the agreement between the different rankings and *PageRank*, *TunkRank*, and "discounted" *PageRank*. In the following section we will discuss the implications of such results.

| PageRank | NodeRanking | "Pruned" PageRank | TunkRank | "Discounted" PageRank | HITS | TwitterRank |
|---|---|---|---|---|---|---|
| * aplusk (actor) | * aplusk (actor) | * cnnbrk (news) | * aplusk (actor) | ryanada_ms (musician) | radioblogger | iphone_app_sale (feed) |
| * cnnbrk (news) | * cnnbrk (news) | * aplusk (actor) | * cnnbrk (news) | aesthetictheory (web designer) | brooksbayne (entrepreneur) | iphone_app_mkup (feed) |
| johncmayer (musician) | johncmayer (musician) | johncmayer (musician) | * stephenfry (actor) | * themandymoore (musician) | stephenkruiser | socialpsych (feed) |
| * stephenfry (actor) | * stephenfry (actor) | * stephenfry (actor) | johncmayer (musician) | astro_127 (astronaut) | twitter_tips | psychnews (feed) |
| * iamdiddy (musician) | * iamdiddy (musician) | * iamdiddy (musician) | * theonion (news satire) | * barbarajwalters (news) | bigrichb | Clareelaine |
| * jimmyfallon (comedian) | * theonion (news satire) | * theonion (news satire) | * jimmyfallon (comedian) | jaygordonmdfaap (doctor) | wbaustin (marketer) | iss_safeguard (feed) |
| * theonion (news satire) | * jimmyfallon (comedian) | * jimmyfallon (comedian) | * ryanseacrest (radio star) | * aplusk (actor) | astronautics | issmontserrat (feed) |
| * ryanseacrest (radio star) | * ryanseacrest (radio star) | * nytimes (news) | * iamdiddy (musician) | * stephenfry (actor) | mattbacak (marketer) | Allenwilk |
| * nytimes (news) | * mashable (news) | * mashable (news) | * katyperry (musician) | fmlteam (blog) | praguebob | Driveorfly |
| * mrskutcher (actress) | * nytimes (news) | * ryanseacrest (radio star) | * mrskutcher (actress) | * cnnbrk (news) | ann_sieg (marketer) | tyneweather (robot) |
| * mashable (news) | * sarahksilverman (comedian) | * sarahksilverman (comedian) | * rustyrockets (comedian) | * breakingnews (news) | jeanettejoy | teesweather (robot) |
| * sarahksilverman (comedian) | * mrskutcher (actress) | * mrskutcher (actress) | * coldplay (musician) | johncmayer (musician) | tmaduri | nyc_tweets (robot) |
| * rustyrockets (comedian) | * rustyrockets (comedian) | * breakingnews (news) | * petewentz (musician) | * nytimes (news) | * joelcomm | phoenix_tweets (robot) |
| * petewentz (musician) | * techcrunch (news) | * techcrunch (news) | * nytimes (news) | * jimmyeatworld (musician) | oliver_turner | o2apps (feed) |
| * katyperry (musician) | * petewentz (musician) | * rustyrockets (comedian) | nprpolitics (news) | webware (news) | andrew303 | apolloapps (feed) |
| * breakingnews (news) | * robcorddry (actor) | * pennjillette (magician) | * danecook (comedian) | crave (news) | startuppro (entrepreneur) | chicago_tweets (suspended) |
| * techcrunch (news) | * snoopdogg (musician) | * robcorddry (actor) | * postsecret (art project) | * google (company) | alohaarleen | rm_extreme (robot) |
| * pennjillette (magician) | * hodgman (actor) | * timoreilly (founder of O'Reilly Media) | * robdyrdek (celebrity) | imeem (music) | orrin_woodward (guru) | thouoaksweather (robot) |
| * robcorddry (actor) | * ev (CEO of Twitter) | * ev (CEO of Twitter) | * google (company) | * soundcloud (music) | upicks (marketer) | sb_weather (robot) |
| nprpolitics (news) | nprpolitics (news) | astro_127 (astronaut) | * pennjillette (magician) | * mpoppel (founder of BNO news) | alicam (guru) | rm_jam (robot) |
| * snoopdogg (musician) | * breakingnews (news) | * hodgman (actor) | * starbucks (company) | Alboebno (BNO news journalist) | oudiantebi (CEO) | isk_g1_1 |
| * hodgman (actor) | * katyperry (musician) | nprpolitics (news) | * chelsealately (show) | felix85 (BNO news contributor) | scotmckay (coach) | sf_tweets (robot) |
| * alyankovic (musician) | * alyankovic (musician) | * petewentz (musician) | * joelmchale (comedian) | Joebrooks | clatko | andyfranks1 |
| * ev (CEO of Twitter) | * mchammer (musician) | * snoopdogg (musician) | * mashable (news) | * astro_mike (astronaut) | brat13 | triciabothwell |
| * mchammer (musician) | * **michaelianblack (comedian)** | * **katyperry (musician)** | * **alyankovic (musician)** | jeffbarr (Amazon evangelist) | robmcnealy (marketer) | rm_club (robot) |
| * michaelianblack (comedian) | * **timoreilly (founder of O'Reilly Media)** | * **mchammer (musician)** | ichcheezburger (humor) | * hollymadison123 (model) | 0boy | rm_harder (robot) |
| * jon_favreau (actor) | * **pennjillette (magician)** | * **alyankovic (musician)** | alancarr (comedian) | sonsofnero (designer) | seanmalarkey (entrepreneur) | isk_g1_10 (robot) |
| * joelmchale (comedian) | * **jon_favreau (actor)** | * **jon_favreau (actor)** | * jason_mraz (musician) | * twitterapi (Twitter API) | suburbview | isk_g1_17 (robot) |
| * starbucks (company) | * **starbucks (company)** | * jack (Co-founder of Twitter) | * marthastewart (entrepreneur) | warped09 (music festival) | coolsi (marketer) | isk_g1_18 (robot) |
| * timoreilly (founder of O'Reilly Media) | * google (company) | * **starbucks (company)** | * markhoppus (musician) | katehavnevik (musician) | caseywright (CEO) | isk_g1_16 (robot) |

**Table 12.** Top-30 users according to different ranking algorithms ordered from left to right in descending order of number of verified accounts atop (marked with an asterisk). A brief description is provided with the user alias. Those users shown in bold appear in the *PageRank* list. Those shaded appear at least in another list.

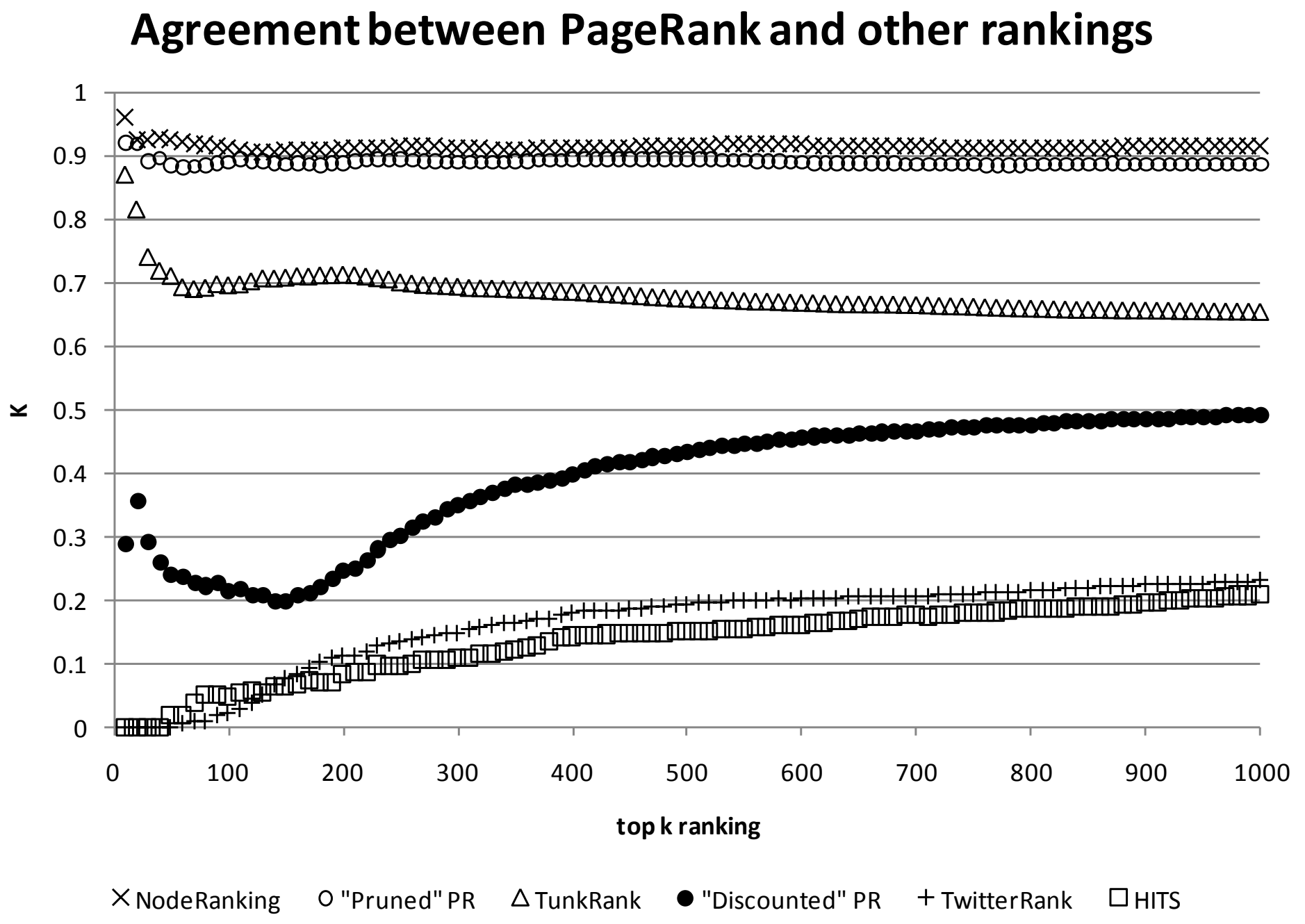

**Figure 7.** Aggrement between *PageRank* and the rest of rankings.

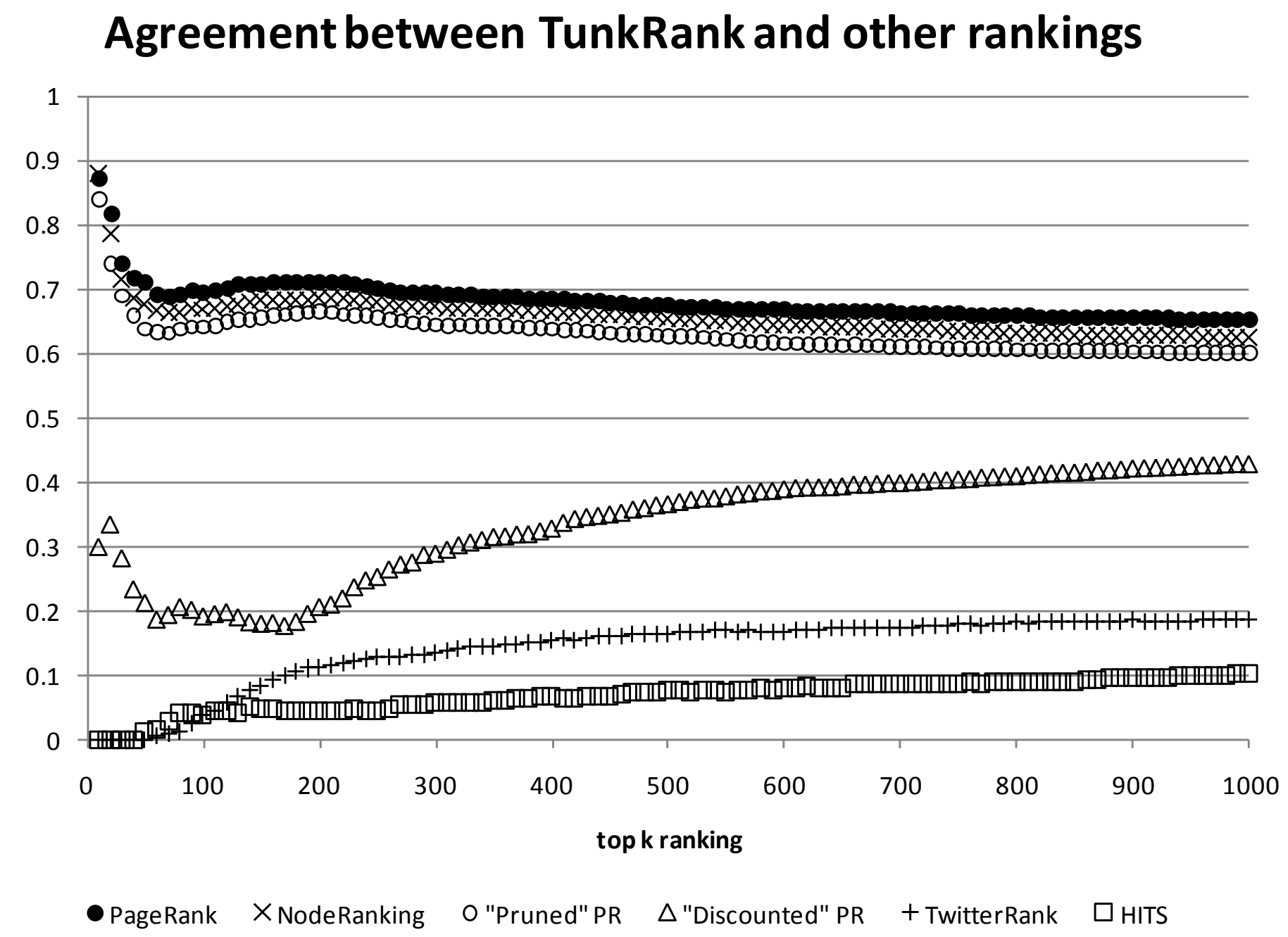

**Figure 8.** Agreement between *TunkRank* and the rest of rankings.

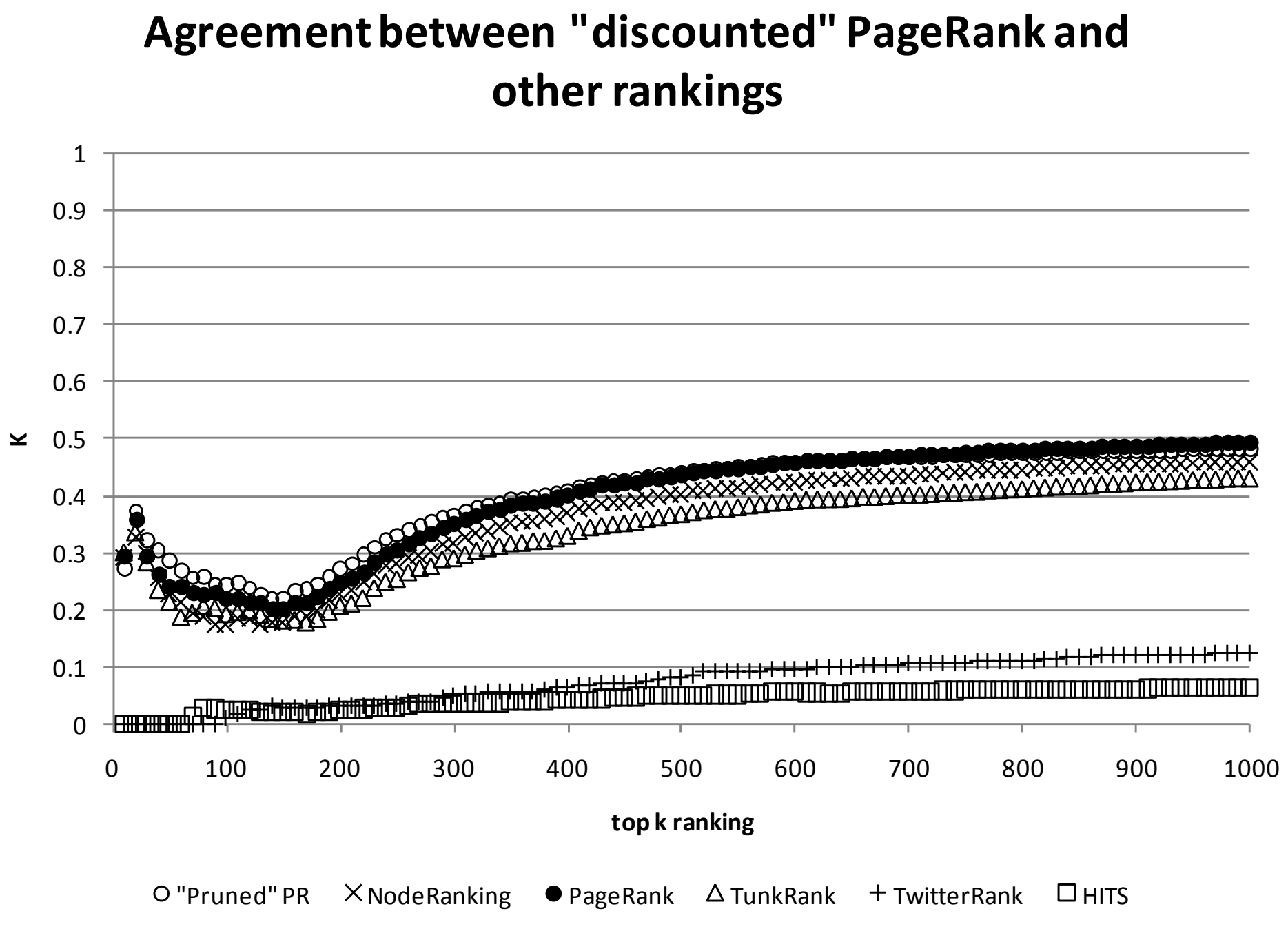

**Figure 9.** Agreement between *PageRank* with paradoxical discounting and the rest of rankings.

# 6. Discussion of results

As we have explained before, our approach to evaluate the available ranking algorithms in the context of social networks was not based on an *a priori* "good" ranking but, instead, their capacity of ranking relevant users (i.e. verified accounts) atop as a whole group while, at the same time, being robust to "gaming" by abusive users, i.e. their "ability" to penalize spammers and aggressive marketers who try to reach better positions by exchanging links instead of providing better content.

Because of the common knowledge about *PageRank* and, additionally, the fact that it seems to be applied by Google to rank Twitter users in their real-time web search engine, we decided to take that method as the baseline against which the rest of techniques should be compared.

The analysis of the results obtained by *PageRank* when applied to the Twitter user graph support our initial concern, that is, one user's *PageRank* is not only a measure of his value within the Twitter ecosystem, but also a consequence of the "tips and tricks" one can employ when establishing relationships within the social network. This is the only plausible explanation for spammers being much better positioned than aggressive marketers when the value of the contents they provide is virtually negligible.

There are two methods which are extremely similar to *PageRank* in terms of ranking abusive and relevant users, and in terms of ranking aggrement: *NodeRanking* and "pruned" *PageRank*. With respect to *NodeRanking*, the similarities are unsurprising given that both NR and PR are highly related. Perhaps *NodeRanking* could reach its full potential with a weighted Twitter graph; however, the way in which such a weighted graph could be inferred from Twitter data (e.g. taking into account the number of mentions or retweets among users, or their content similarity) is out of the scope of this paper. In this regard, however, we must note that both methods slightly underperform *PageRank*: on one hand, they promote to top positions a slightly lesser number of verified accounts than *PageRank* and, on conversely, they promote a slightly larger number of abusive users than *PageRank*.

The similarity between the results obtained by *PageRank* and *PageRank* applied to the "pruned" Twitter graph are somewhat expected; however, they deserve a deeper analysis because they support of the points of this author. Let us remember that the "pruned" graph was obtained by removing those users (and their in- and out-links) with zero de-weighting which, in turn, was computed taking into account reciprocal links between users. One of the arguments of this author is that discounting reciprocal links is a pretty fine way to separate users contributing to the ecosystem, from those with little or no value at all. The results obtained with the total Twitter graph and the "pruned" graph are virtually the same and, thus, we can take that as supportive of the fact that most reciprocal links are not legitimate but, instead, merely an aim to reach larger audiences.

In contrast, there are two methods which greatly differ not only from *PageRank* but also from the other techniques, namely *HITS* and *TwitterRank*. Each of them exhibits different problems when applied to the Twitter graph.

*HITS* underperforms *PageRank* with respect to both verified accounts and abusive users. As it can be seen in Table 12 the top list produced by *HITS* is plagued by mostly irrelevant users (at least compared with the top lists produced by the other methods) which can be also appreciated in Figure 3. Moreover, if we checked the number of followers and followees for those users we could see that the ratio for most of them is close to 1 and, in fact, most of them have got a large number of reciprocal links.

In fact, because of the very nature of *HITS*, this algorithm is virtually inoperative when confronted with a relatively small number of users weaving a tight network of reciprocal connections. Let's remember that many spammers and marketers tend to massive follow other users in order to gain a follow-back link. Because of this, when computing hub scores those users' values tend to grow very fast; then, those hub scores are used to compute authority scores for their followees (which are mostly spammers and following them back). It is clear that after several iterations those users with lots of reciprocal links earn an undeserved amount of authority. Hence, the *HITS* algorithm is not advisable at all to rank users within social networks since it is clear that *HITS* is not robust to follow-spam.

The results achieved by *TwitterRank* were very disappointing. Conceptually, it is a very appealing method: it provides ways to incorporate both content similarity measures and transition probabilities into the ranking. Some way, however, these appealing ideas seem to fail: as it can be seen from the top list, most of the users are feeds and robots, many of them highly related (even with strikingly similar names). To be fair it must be said that modifying a topic-sensitive method to operate globally is, perhaps, pushing too hard the technique. However, given that even the simplified version implemented for this research (using cosine similarity instead of LSA) is (1) much more computationally expensive than the rest of methods surveyed, and (2) it requires much more data (namely, the tweets) to obtain the rankings, it seems not at all recommendable, especially when other available methods (e.g. *TunkRank*) are faster and provide much better results (at least when applied globally to the complete user graph).

Lastly, there are one method clearly outperforming *PageRank* with respect to both ranking atop verified accounts and penalizing abusive users: *TunkRank*. It is certainly similar to *PageRank* but it makes a much better job spotting relevant users and also when confronted with "cheating": aggressive marketers are almost indistinguishable from common users –which is, of course, desirable; and spammers just manage to grab a much smaller amount of the global available prestige and reach lower positions –although they still manage to be better positioned than average users. In addition to that, the ranking induced by *TunkRank* certainly agrees with that of *PageRank*, especially at the very top of the list, and, moreover, it is able to put a larger number of verified accounts at top ranks. Thus, *TunkRank* is a highly recommendable ranking method to apply to social networks: it is simple, it prominently ranks relevant users, and severely penalizes spammers when compared to *PageRank*.

With regards to the variation to *PageRank* described by this author, "discounted" *PageRank*, the results are not highly conclusive.

On one hand, performance against the proposed benchmarks is pretty good. With regards to demoting abusive users, it seems to outperform *PageRank* –and even *TunkRank*– because the amount of prestige grabbed by such users is smaller and their rankings lower than when applying standard *PageRank*. With regards to spotting verified accounts, as it is shown in Figure 4, it outperforms *PageRank* (although underperforms *TunkRank*).

Nevertheless, it has two issues which deserve further research. On one hand, the induced ranking could be labeled as "elitist" because most of the users –about 70%– tie for the last position. Certainly, this is unsurprising given that 16% of the users in the graph have got a zero de-weighting factor what we interpret as their contributions being "worthless" for the network as a whole. Moreover, such results are consistent with the well-known participation inequality [41], and with a recent study revealing that 75% of the users just publish a tweet every 9 days, and 25% of the users do not tweet at all [25]. Hence, this could be considered a minor issue.

On another hand, "discounted" *PageRank* exhibits a fairly distinctive curve (see Figure 9) when comparing its agreement with other rankings –obviating the underperforming *HITS* and *TwitterRank*. The agreement is much lower than, for instance, that found between *PageRank* and *TunkRank*, but the most striking behavior is the local maximum at the top positions, followed by a relatively large trough, to eventually stabilize. We found several lesser-known users at top ranks and, after studying them, we concluded that most of them have one or more "famous" followers who, in many cases, they manage to outrank. We have denoted this as the "giant shoulders" effect and it explains not only the trough at the head of the list but the smaller agreement for the rest of the ranking: many of the top users from *PageRank* or *TunkRank* are a little behind of lesser-known users they are following. This is aesthetically displeasing, at least, and the effect it can exert in the applications of the ranking is still to be explored. Nevertheless, tackling with this and the former issue is left for future research.

# 7. Implications, conclusions, and future work

This study makes four main contributions. First, when applying graph prestige algorithms to social networks, ranking is not only a measure of a user's value but also the result of "gaming" the algorithm by means of relationship links. The fact that spammers –who contribute no valuable content– are consistently better positioned than marketers –who contribute somewhat valuable information– no matter the method employed supports this assert.

Second, evaluating ranking should not be a point in itself; it should, instead, be evaluated within an objective context. To compare the performance of different algorithms two metrics are to be applied: First, relevant users should reach top ranks as a whole without regards to their internal ordering. Secondly, abusive users should not reach undeserved rankings.

Third, *TunkRank* is an obvious candidate to rank users in social networks. Although highly related to *PageRank*, *TunkRank* outperforms it with respect to both spotting relevant users in top positions, and penalizing abusive users. In addition to that, it is simple to implement and computationally cheap –at least as cheap as *PageRank*.

And fourth, reciprocal linking in social networks is, in many cases, a sign of follow spam instead of mutual interest among legitimate users. This is supported by two experimental findings:

- First, by applying *PageRank* to a Twitter graph where users with zero de-weighting were removed we achieved virtually the same results than when applying the same algorithm to the whole graph. In other words, most of the reciprocal links do not provide any information at all.
- Second, when incorporating to *PageRank* a way to de-weight influence on the basis of reciprocal links spammers are highly demoted while a higher number of relevant users are found at top positions. In other words, although most of the reciprocal links provide little information this is not applicable to all of them and, hence, they should not be simply removed but accounted for as shown in Section 3.4.

This study opens several lines of research. First, the rankings induced by the different methods should be analyzed in other contexts, for instance, as a way to rank content providers in order to find relevant information within a social network. Second, *TunkRank* is not immune to manipulation and, thus, its vulnerabilities should be thoroughly studied (e.g. Sybil attacks could be a starting point). And third, a deeper analysis of the role of nepotistic links, in general, and the "discounted ratio" described in this paper, in particular, is needed.

# 8. Acknowledgements

The author would like to thank David J. Brenes, Miguel Fernández, Fernando Zapico, and Diego Guerra for their help during the Twitter dataset collection. He is also in debt with Miguel Fernández for comments on an early draft of this paper. This work was partially financed by grant UNOV-09-RENOV-MB-2 from the University of Oviedo. Finally, the author would like to thank the anonymous reviewers for their constructive suggestions to improve this paper, especially regarding the use of ground truth data about relevant users.

# Appendix. In-depth Analysis of the Twitter Dataset

The study described in this paper relied on a Twitter dataset collected by the author along 2009. The dataset is composed of two different parts: a collection of 27.9 million English tweets, and a user graph comprising 1.8 million users and 134 million connections. In this Appendix we provide an in-depth analysis of such dataset: we describe both the network characteristics and several demographical features of the users in the network.

Table A-1 shows some statistics describing the collected graph and compares it with the graph previously built by Java *et al.* [29], and with the whole Twitter crawl by Kwak *et al.* [31]. Some of the values for those graphs are fairly similar –or at least comparable– while those with bigger differences can be attributed, in all probability, to the Twitter growth in the last two years, in addition to sampling artifacts. In fact, when comparing the graphs by this author and Java *et al.*, the increase in the average degree, the size of both the largest WCC and SCC, and the clustering coefficient is consistent with a growth in the number of users together with a larger number of connections between them.

| Property | Twitter sample (2009) | Twitter graph (2009) [31] | Twitter in 2007 [29] |
|---|---|---|---|
| Total nodes | 1,804,131 | 41.7M | 87,897 |
| Total links | 134,500,669 | 1,047M | 829,247 |
| Average Degree | 74.55 | 25.11 | 18.86 |
| Indegree Slope | -1.33 | -2.276 | -2.4 |
| Outdegree Slope | -1.516 | N/A | -2.4 |
| Degree Correlation | 0.490 | N/A | 0.59 |
| Diameter | 6 | 4.8 (effective diameter [50]) | 6 |
| Largest WCC size | 1,800,132 (99.78%) | N/A | 81,769 (93.03%) |
| Largest SCC size | 1,688,395 (93.58%) | N/A | 42,900 (48.81%) |
| Clustering coefficient | 0.151 | N/A | 0.106 |
| Reciprocity | 0.48 | N/A | 0.58 |

**Table A-1.** Network properties for three different Twitter graph crawls. From left to right, the graph collected by this author, a whole complete graph crawled in 2009 [31], and a subgraph collected in 2007 [29].

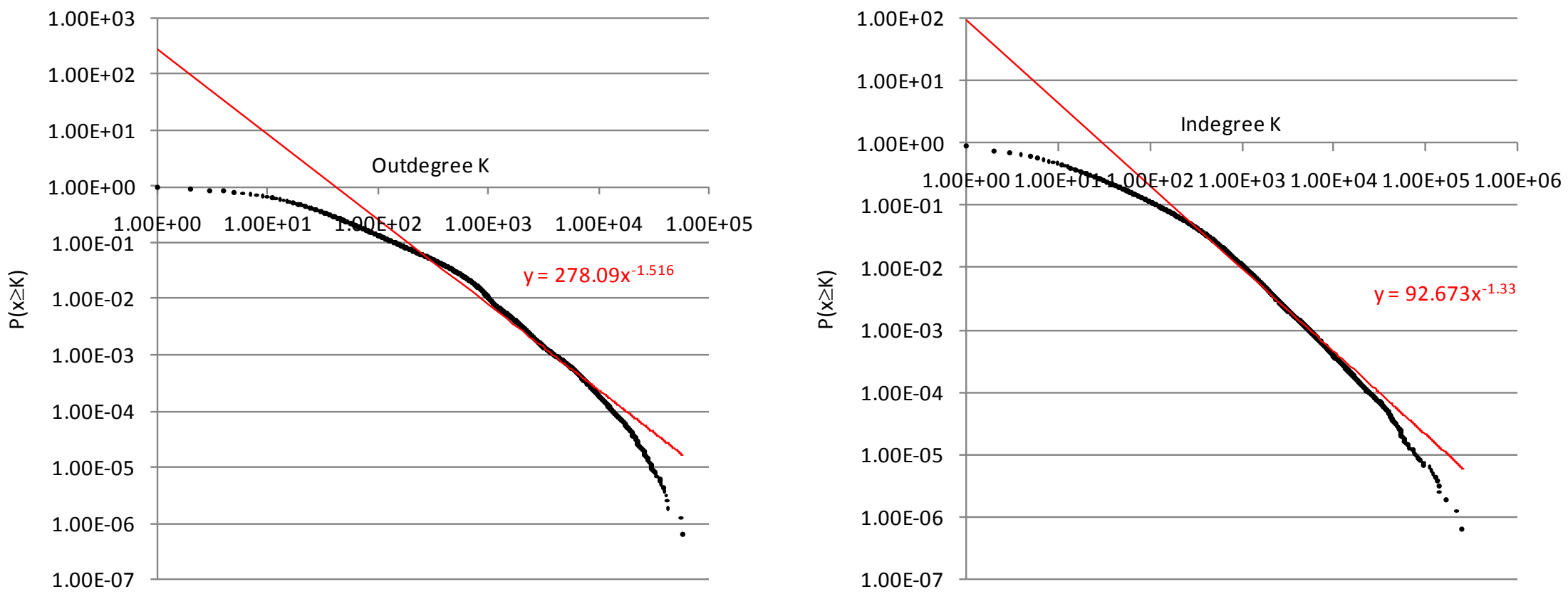

**Figure A-1.** Outdegree and indegree distributions in the Twitter graph. Both exhibit a power law exponent (-1.516 for the outdegree and -1.33 for the indegree). Surprisingly, Kwak *et al.* argue that the follower distribution for the whole Twitter graph does not follow a power law [31].

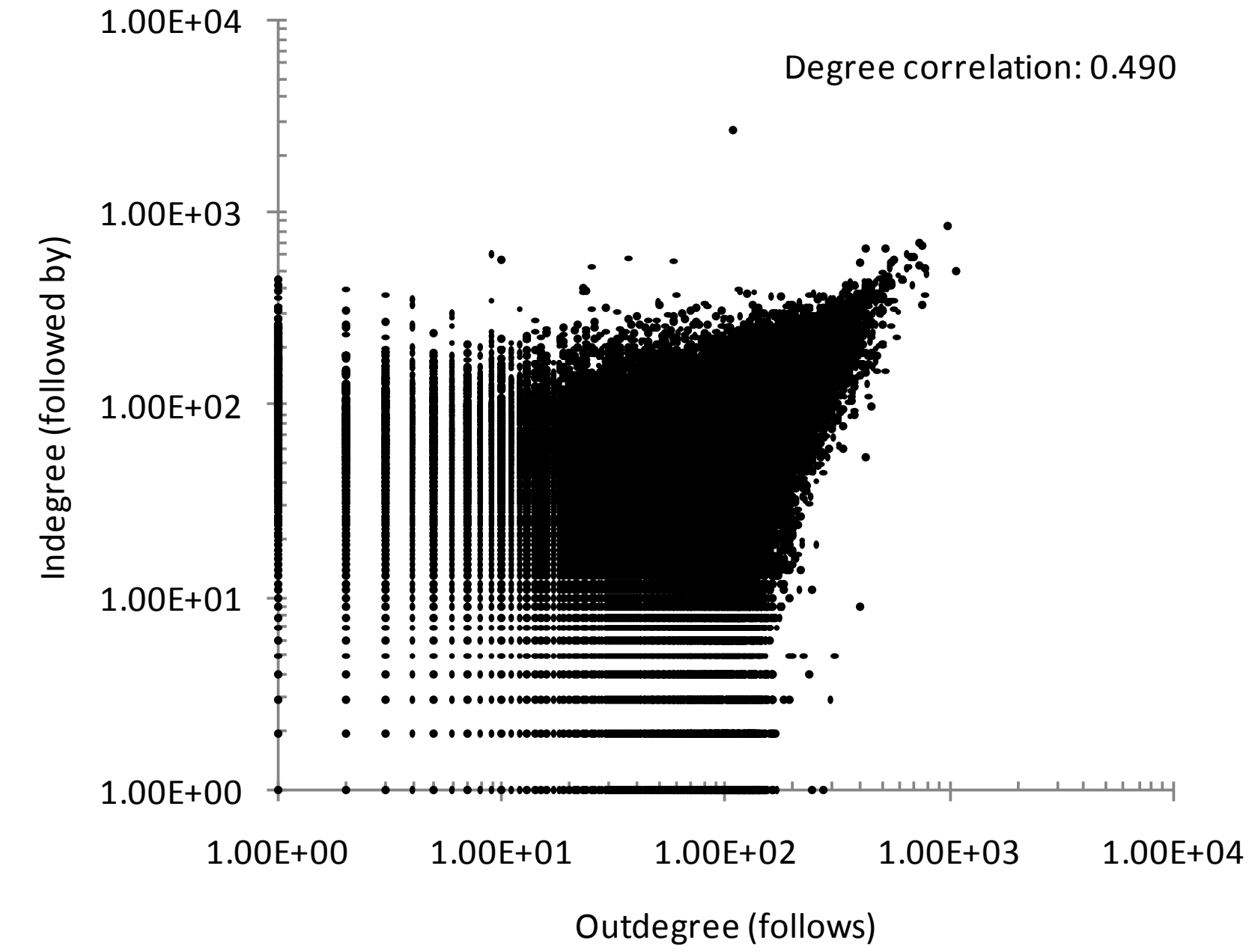

**Figure A-2.** Scatter plot showing the correlation between in- and out-links in the Twitter user graph. The 0.49 correlation is similar, although somewhat smaller than the reported in [29].

During the graph collection profile information was obtained for each user, namely the user's name, location, biography and website (see Table A-2 for an extract of such profiles). Such information was employed to determine the geographical location, gender, and age of the users.

It must be noted that locations are nothing but free text and, thus, it's up to the user providing a sensible location (e.g. *London* or *NYC*) or something mostly irrelevant (e.g. *at home* or *in the office*). We processed the available locations (62.31% of the users provide such a string) and tried to match them to geographic coordinates by means of a geocoding service[20]. Eventually, 50.36% of the original profiles provided a location string suitable to be matched to actual coordinates.

Given the noisiness of locations, one could argue that some, even many, of the obtained coordinates could be wrong. Nevertheless, as can be seen in the map in Figure A-3 most of the locations must be necessarily correct: users from English speaking countries are majority (the USA and the British Isles are specially prominent); Canada, Australia, New Zealand, Jamaica, Puerto Rico, Netherlands, central Europe and Israel have also a major presence in the sample. Finally, there exist pockets of English-tweeting users in virtually every country but concentrated, as expected, in major global cities (e.g. Paris, Tokyo, Madrid, Seoul, Buenos Aires or São Paulo). On the other hand, the distribution within English speaking countries faithfully corresponds with their population density. So, in short, it seems safe to claim that half of Twitter users provide an accurate geographical location.

The name and biography fields were in turn employed to infer some demographic features about the sampled users, namely gender and age. To determine the gender we parted from the *"Frequently Occurring First Names and Surnames From the 1990 Census"*[21]; those data files provide 88,799 surnames, 4,275 female first names and 1,219 male first names. We assumed that any user name starting with a first name and ending with a last name from the census was a valid personal name. Certainly, many people provide aliases, just their first name, or their names and/or surnames are not frequent enough to appear in the U.S. Census data; however, we think that this approach is the best for the sake of higher precision.

With regards to those first names appearing in both male and female data files (e.g. *Alexis*, *Charlie*, or *Dominique*) we assign gender according to the frequency of appearance provided the difference was high enough. In this regards, *Alexis* and *Dominique* were always considered female names while *Charlie* was considered a male name. Of course, this is an oversimplification which, certainly, could be improved by taking into account the data in the biography field but we considered that, for the descriptive purposes of this section, it is good enough.

[20] http://developer.yahoo.com/maps/rest/V1/geocode.html

[21] http://www.census.gov/genealogy/names/

To support that claim some anecdotal evidence can be provided. First, there exists an almost perfect positive correlation between the last name distribution in the U.S. 1990 Census and within the Twitter users (0.9701). The correlation regarding first name lists is smaller but still positive (0.6355 for female names and 0.6356 for male names). Arguably, this can be due to a major presence of young users among *twitterers*. As it can be seen in Table A-3 just one female name appear in both top-10 lists (*Jennifer*) while three are common for male names (*James*, *John* and *Michael*). Both situations seem consistent with the fact that given names have relatively fast turnovers (well under a decade), in particular, female names (cf. [35]).

In addition to this, we computed the most distinctive terms in both male and female biographies by means of a likelihood-ratio test in a way analog to that of [29]. Among the top-10 words for females were *mom*, *girl*, *wife* and *mother*, while *husband*, *guy*, *father*, *dad* and *man* appeared at the top of the list for male users; Table A-4 provides a more exhaustive list.

Hence, it seems that our method to assign gender to Twitter users is reasonably accurate and, thus, it can help to provide a picture of the demographics of these users. About 650,000 users provide a personal name (both first and last name appearing the 1990 Census data files), accounting for 36.46% of the users in the graph, from which 58.61% were men and 41.39% women.

| | | | | |
|---|---|---|---|---|
| Bambaloo | Emma Bullen | London | | |
| becky_mallery | becky mallery | Kent (but second home london!) | MSc Positive Psychology Student, Author and Coach | |
| Burkjackson | burkjackson | iPhone: 46.181351,-123.818344 | creative, thinker, father, dreamer... | http://www.Jackson5Home.com/ |
| natekoechley | Nate Koechley | Home | I split time between online & San Francisco. One wife, one daughter, two cats, three vices & four eyes. Now: Outspark's VP of UX. Previously Yahoo! & YUI. | http://nate.koechley.com/blog |
| pete_watson | pete_watson | England - home of the uprising | | |
| scottisafool | Scott Lovegrove | 50.69052061,-1.93774726 | Software tester for a telecomms company in South England | http://scottisafool.spaces.live.com/ |
| ShaLaylaj | ShaLayla J. Simmons | Too far from home | I've actually been hailed as probably the most significant woman who's ever existed (I was hoping for Jet Beauty of the Week, but I can make this work). | |
| Swpatrick | Patrick Littlemore | London, South & West | London Career Estate Agent (Lettings). Married and father of one amazing angel. An Australian who loves London! | http://tiny.cc/Tenants |
| TheRealSani | Sandra S. | Home, sweet home | Hi, I'm Sani and i believe in GOD. I love peace, butterflys, Retro, Mick Jagger and how my life's going... | |
| ZimHilton | Hilton Barbour | Misty London UK | Networker. Loves travelling - love my 2 daughters more. Perpetually curious. | http://hiltonbarbour.com |

**Table A-2.** An extract of the user profiles contained in the dataset. As it can be seen these profiles comprised the screen name, the user name, location, short biography and website. All of the fields, except for the first two, are optional.

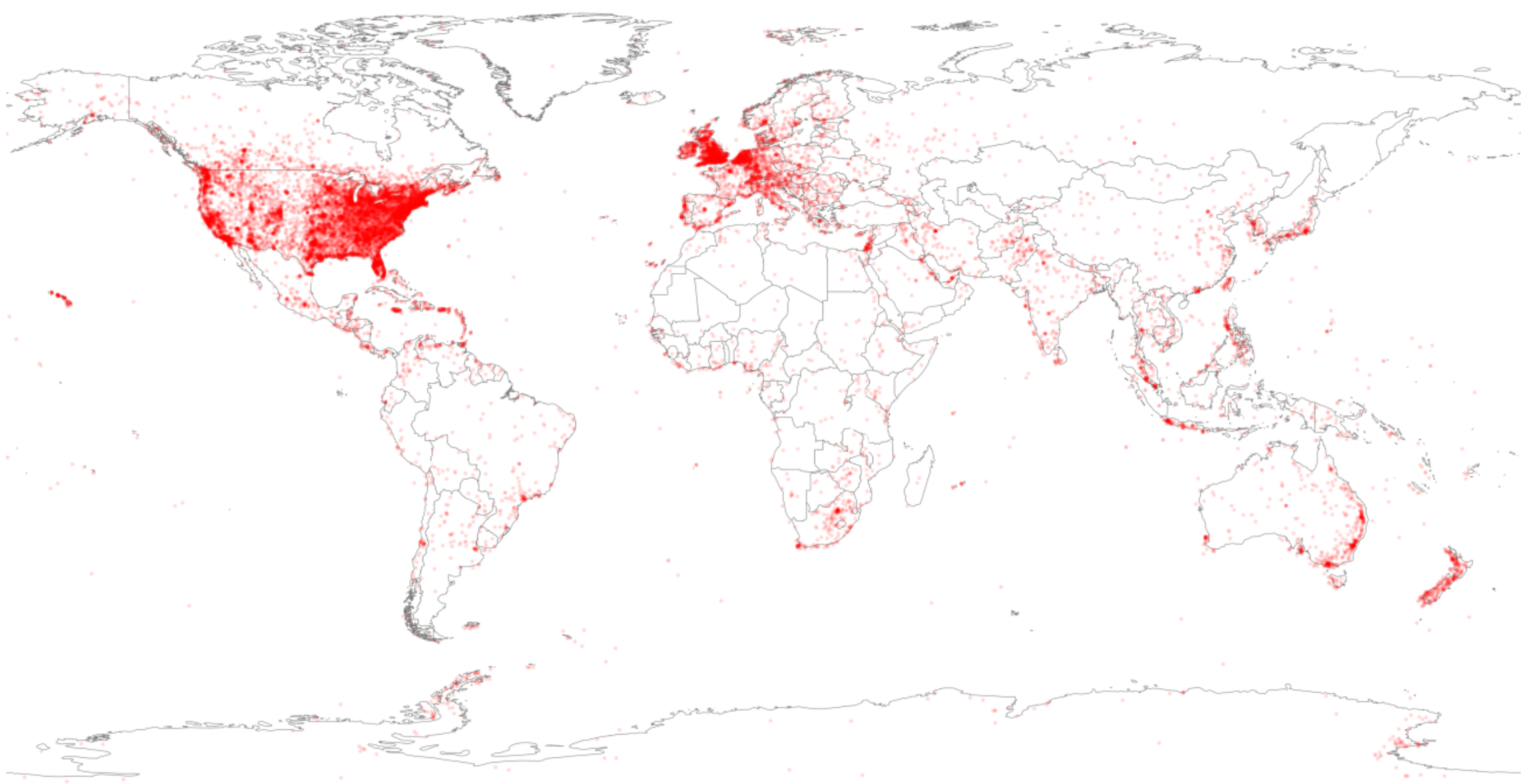

**Figure A-3.** Map showing the global distribution of users in the dataset. Let's remember that it only contains tweets written in English and, thus, English-speaking countries should concentrate most of the users.

| 10 most frequent last names in Twitter | 10 most frequent first female names in Twitter | 10 most frequent first male names in Twitter | 10 most frequent last names in the U.S. 1990 Census | 10 most frequent first female names in the U.S. 1990 Census | 10 most frequent first male names in the U.S. 1990 Census |
|---|---|---|---|---|---|
| **Smith** | Sarah | Chris | **Smith** | Mary | **James** |
| **Johnson** | **Jennifer** | **Michael** | **Johnson** | Patricia | **John** |
| **Jones** | Amanda | **John** | **Williams** | Linda | Robert |
| **Brown** | Michelle | **James** | **Jones** | Barbara | **Michael** |
| **Williams** | Amy | Mark | **Brown** | Elizabeth | William |
| **Miller** | Stephanie | Matt | **Davis** | **Jennifer** | David |
| **Davis** | Rachel | **David** | **Miller** | Maria | Richard |
| Lee | Heather | Mike | **Wilson** | Susan | Charles |
| **Wilson** | Katie | Paul | Moore | Margaret | Joseph |
| **Taylor** | Jessica | Andrew | **Taylor** | Dorothy | Thomas |

**Table A-3.** List showing the top-10 first and last names in both Twitter and the U.S. 1990 Census. Names appearing in both lists are shown in bold.

| Top 50 terms appearing in female biographies | | | Top 50 terms appearing in male biographies | | |
|---|---|---|---|---|---|
| love | happy | people | **husband** | media | radio |
| life | friend | art | **guy** | producer | manager |
| **mom** | music | beautiful | web | internet | church |
| **girl** | world | boys | **father** | marketing | digital |
| **wife** | **woman** | single | developer | photographer | **actor** |
| **mother** | **mum** | amazing | geek | fan | programmer |
| friends | married | **wife mother** | **dad** | **husband father** | web developer |
| lover | **mommy** | **actress** | **man** | pastor | guitar |
| loves | **sister** | kids | designer | video | ceo |
| fun | person | books | musician | gamer | enthusiast |
| live | dance | hi | sports | computer | **dude** |
| family | heart | love music | technology | business | play |
| crazy | home | lol | tech | founder | dj |
| loving | little | dog | director | consultant | online |
| laugh | want | chick | software | design | beer |
| fashion | know | teacher | entrepreneur | player | development |
| **daughter** | reading | | engineer | games | |

**Table A-4.** Top 50 more distinctive terms for female (on the left) and male (on the right) users in Twitter. Those terms which have a clear associated gender are shown in bold. The rest of terms do not have any prior gender but some patterns arise: female users tend to describe their family life, while male users tend to describe their occupations.

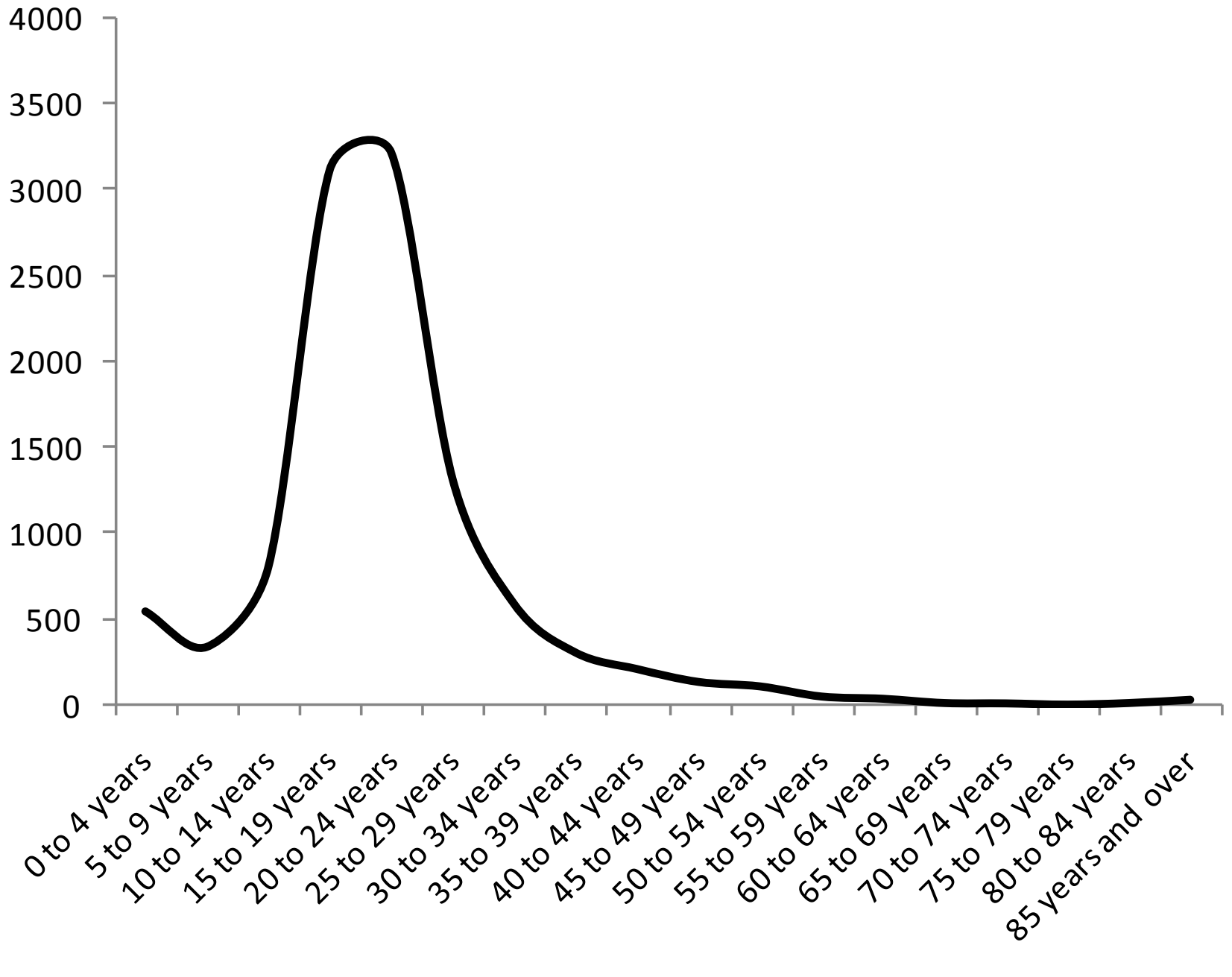

**Figure A-4.** Age distribution of Twitter users.

In addition to gender, we also tried to determine the age of the users. To that end we relied on the biographies appearing in the profiles, and employed simple patterns (i.e. looking for *"years old"* or *"year-old"* preceded by a number or a numeral). This is also a crude approach, and prone to some mistakes (consider for instance the phrase *"proud mother of 6 years old boy"*) but, all in all, we consider the results highly plausible. Only 10,915 users provide an identifiable "age" in their biographies, the average is 21.67 years with standard deviation of 11.086 years. The distribution is clearly not normal, and users between 15 and 29 years account for 70% of the population. Such a bias toward young adults is, however, expected given the social nature of Twitter, and consistent with the previous discussion about first names.

As we did with male vs. female users, we also obtained distinctive terms from the users' biographies within each age range. Such terms do provide some clues on the confidence one can place on the aforementioned method to determine the age of Twitter users. Table A-5 shows how the most distinctive terms almost perfectly fit the usual age stereotypes in the U.S. and the UK. For instance, 10 to 14 year-old users are fans of *Twilight*, *Jonas Brothers* and *Miley Cyrus* while users from 15 to 24 years tend to mention *school*, *college* and *university*. Users from 25 to 59 years are commonly married, with kids being especially prominent from 30 onwards. Grandparenthood appears as early as 45-49 years, but is much more common from 55 onwards. Lastly, from 60 to 74 years old, retirement does appear.

Needless to say, there are mistakes: as it was expected, users from 0 to 9 years old are not in fact that age but parents with little children, instead. With regards to the age range of 85 years and more, it seems that most of the users are not indeed that age given the lack of terms appearing in the immediately prior ranges, besides the great heterogeneity of the distinctive terms for that age. If those age ranges are dropped, then the average age is 21.13 years with standard deviation of 9.08 years.

Finally, by combining age and gender it is possible to produce a population pyramid for Twitter (see Figure A-5). Only 4,295 users have both gender and age information and, thus, it's not entirely representative of Twitter users as a whole. However, we can extract some knowledge from it: the current prominence of male users against female users is expected to gradually change towards parity, even, surpassing the number of male users; as it can be seen, female users in the 10-14 and 15-19 ranges (the so-called *digital natives*) clearly surpass the number of male users.

| | |
|---|---|
| 0 to 4 years | **mom**, **son**, **daughter**, **mother**, **boy**, **wife**, home, stay, little, **married**, beautiful, **father**, **home mom**, **twins** |
| 5 to 9 years | **mom**, **son**, **daughter**, **mother**, **twins**, **boy**, **married**, **boys**, **twin**, **wife**, **father**, **dad**, **home**, marketing, **husband** |
| 10 to 14 years | love, girl, name, **twilight**, hey, **jonas brothers**, boy, follow, twitter, **miley cyrus**, grade, **fan**, live, demi, hi |
| 15 to 19 years | music, girl, love, friends, name, **school**, **student**, hey, follow, hi, know, **college**, loves, **jonas brothers** |
| 20 to 24 years | **college student**, life, **university**, **major**, world, **studying**, design, music, working, fun, **graduate**, living, trying |
| 25 to 29 years | **married**, living, work, **wife**, working, male, geek, **mother**, guy, **kids**, writer, teacher, **husband**, manager |
| 30 to 34 years | **married**, **kids**, **mother**, **husband**, **father**, male, lover, single, **children**, beautiful, work, **wife**, female, living |
| 35 to 39 years | **married**, male, **boys**, **children**, single, **mom**, good, **father**, **kids**, **mother**, radio, business, **mum**, guy |
| 40 to 44 years | **married**, single, man, **kids**, **father**, **children**, **mother**, living, woman, **wife**, **mom**, wonderful, male, dogs |
| 45 to 49 years | **grandmother**, **boys**, **wife**, **married**, **kids**, **wonderful boys**, **children**, marketing, **father**, **mother**, **sons**, happily |
| 50 to 54 years | **married**, woman, adult, male, single, **children**, people, **daughters**, man, **father**, cooking, **grandchildren** |
| 55 to 59 years | **married**, cats, **children**, internet, **grandchildren**, hanging, **ex**, man, spiders, surgery, **widow**, male, gardening |
| 60 to 64 years | **retired**, **grandchildren**, south, life, archaeology, flowers, **traveller**, **enjoying**, electro, goals, lawyer, physics |
| 65 to 69 years | man, diving, body, time, **retired**, country, lives, technology, marketing, business, likes, single, school |
| 70 to 74 years | **grandfather**, blogs, running, **retired**, tech, computer, blogger, man, female |
| 75 to 79 years | **granddaughters**, dog, **grandchildren**, stories, bike, grown, tennis, life, living, almost, **children**, play |
| 80 to 84 years | help, **pension**, **retire**, listener, independent, look, money, others, yeah, company, movies, love |
| ≥85 years | newspaper, home, cats, care, aquarium, satisfied, port, bones, river, trades, blogs, books, lol, reader, tea |

**Table A-5.** Distinctive terms from the users' biographies for each age range. Clearly stereotypical terms are shown in bold.

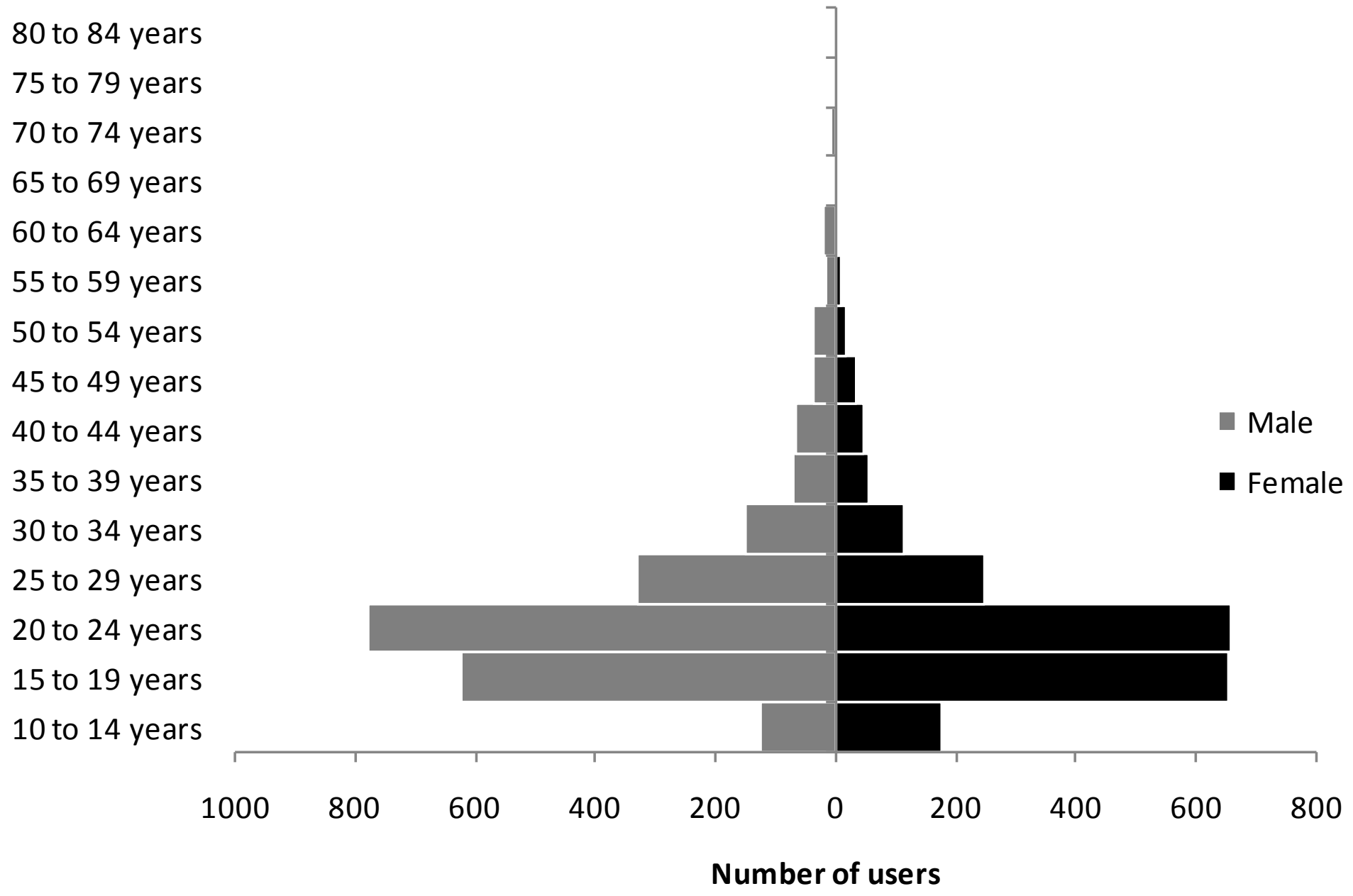

**Figure A-5.** Population pyramid for Twitter built from the biographies of the 4,295 users who provided both age and gender information.

To sum up, our dataset consists of two different collections: a series of tweets, and the network of users who published them. The tweets collection comprises 27.9 million English tweets posted between January 26, 2009 and August 31, 2009. The user graph consists of 1.8 million users with 134 million connections. 50% of the users are geolocatable, and most of them reside, as expected, in English speaking countries. It was possible to determine gender for one third of the users; of these, 59% were men and 41% women. A relatively small number of users provide age information. Taken them as sample, the average age in Twitter is 21 years and users between 15 and 29 years account for 70% of the total. By examining both age and gender, it seems that the male predominance in Twitter has its days numbered because female surpass male users among the youngsters (10 to 19 years old).